\documentclass[12pt]{article}

\usepackage{graphicx}
\usepackage{dcolumn}
\usepackage{bm}
\usepackage[            
pdftex,                 
pdfborder= 0 0 0,
a4paper,                
colorlinks=true,        
citecolor=green,        
linkcolor=cyan,         
menucolor=blue,         
urlcolor=magenta,       
bookmarksopen=true
]{hyperref}

\usepackage{amsmath}
\usepackage{amssymb}

\newcommand{\PNAS}{Proceedings of the National Academy of Sciences 
                   of the United States of America}
\newcommand{\Ps}{\P^A_{\mbox{ses}}}
\newcommand{\Psz}{\P^{A,0}_{\mbox{ses}}}
\newcommand{\Es}{\E^A_{\mbox{ses}}}
\newcommand{\Esz}{\E^{A,0}_{\mbox{ses}}}

\newcommand{\Rs}{R_{\mbox{ses}}}

\newcommand{\z}{Z\kern-0.45emZ}
\newcommand{\vi}{I\kern-0.3emB}
\newcommand{\1}{I\kern-0.3emI}

\newcommand{\e}{I\kern-0.3emE}
\newcommand{\E}{I\kern-0.3emE}
\newcommand{\ind}{I\kern-0.3emI}
\newcommand{\p}{I\kern-0.3emP}
\renewcommand{\P}{I\kern-0.3emP}

\newcommand{\be}{\begin{equation}}
\newcommand{\ee}{\end{equation}}

\begin{document}

\title{Two-level Fisher-Wright framework with selection and migration: An approach to studying evolution in group 
structured populations}

\author{Roberto H. Schonmann$^{1,2}$, Renato Vicente$^2$, Nestor Caticha$^3$}

\date{\today}

\maketitle 

\begin{center}
1. Dept. of Mathematics, University of California at Los Angeles, 
CA 90095, USA \\
2. Dept. of Applied Mathematics, Instituto de Matem\'atica e Estat{\'\i}stica, 
Universidade de S\~ao Paulo, 05508-090, S\~ao Paulo-SP, Brazil\\
3. Dep. de F{\'\i}sica Geral, Instituto de F{\'\i}sica, 
Universidade de S\~ao Paulo, CP 66318, 05315-970, S\~ao Paulo-SP, Brazil
\end{center}

\vspace{1cm}
\noindent {\bf Key words and phrases:}{ Natural selection; Fisher-Wright model;
population genetics; evolutionary game theory;
trait-group framework; altruism; cooperation;
kin selection; group selection; Price equation; Hamilton's rule; relatedness;
neighbor-modulated fitness;
iterated public goods game; generalized tit-for-tat strategies;
threshold models; strong and weak selection;
multitype branching processes; Perron-Frobenius eigenvalue and eigenvector;
viability or survivability criterion; survival mechanism; Wright's infinite islands 
model.}

\vspace{3.0cm}
\noindent{\bf Acknowledgements:} {R.H.S. is glad to thank Rob Boyd for many hours of stimulating and informative conversations 
on the subjects in this paper. He also 
warmly thanks Marek Biskup for finding a derivation of 
(\ref{mKDexp}) in  a special case; that
result motivated us to engage in the work that eventually resulted in Section 5 of
this paper. R.H.S. is also grateful to Clark Barrett, Maciek Chudek, Daniel Fessler, Sarah Mathew and
Karthik Panchanathan for nice conversations and feedback on various aspects of this project and related subjects. This project was 
partially supported by CNPq, under grant 480476/2009-8.}

\vspace{1.0cm}
\noindent{\bf Emails: }{\href{mailto:rhs@math.ucla.edu}{rhs@math.ucla.edu}, 
\href{mailto:rvicente@ime.usp.br}{rvicente@ime.usp.br}, 
\href{mailto:nestor@if.usp.br}{nestor@if.usp.br} }

\newpage

\begin{abstract}
A framework for the mathematical modeling of evolution in group structured
populations is introduced.
The population is divided into a fixed large number of groups of fixed size.
From generation to generation, new groups are formed that descend from
previous groups, through a two-level Fisher-Wright process, with selection
between groups and within groups and with migration between groups at rate $m$.
When m=1, the framework reduces to the often used trait-group
framework, so that our setting can be seen as an extension of that approach.
Therefore our framework is sufficiently flexible to allow the analysis of many
previously introduced models in which altruists and non-altruists compete,
and provides new insights into these models. We focus on the situation in 
which initially there is a single altruistic allele, in the population, and
no further mutations occur. The main questions are conditions for the viability
of that altruistic allele to spread, and the fashion in which it spreads when it does.
Because our results and methods are mathematically rigorous, we 
see them as 
shedding light on various controversial issues in this field, including the role
of Hamilton's rule, and of the Price equation, the relevance of linearity in fitness 
functions, the need to only consider 
pairwise interactions, or weak selection, etc. 
In the current paper we analyze the early stages of the evolution, 
during which the number of altruists is small compared to the size of the 
population. We show that during this stage the evolution is well described by a multitype 
branching process. The driving matrix for this process can be readily obtained, reducing
the problem of determining when the altruistic gene is viable to a comparison 
between the leading eigenvalue (Perron-Frobenius eigenvalue) of that matrix, and the fitness 
of the non-altruists before the altruistic gene appeared.
This leads to a generalization of Hamilton's condition for the viability of 
a mutant gene. That generalized viability condition can be interpreted in an appropriate neighbor 
modulated fitness sense, providing a gene's eye view of the generalized rule.  
Our generalized Hamilton rule reduces to the traditional one for 
public goods games, and more generally under the condition of linearity of 
the fitness of each carrier of the gene A as a function of the number of copies 
of that gene in the same group. 
Our analysis also suggests a broadly applicable criterion, that 
we make explicit, for the viability of a mutant gene, in a more general setting.   
Our generalized Hamilton condition simplifies considerably when selection is weak,
and further when groups are large. We analyze a significant number of examples, 
and observe that the altruistic gene can spread under relatively low
levels of relatedness in the groups, corresponding to relatively high levels of
migration. This happens, for instance, when the fitness of individuals
is affected by  
repeated activities in their groups, and the altruistic 
mutant gene promotes cooperation 
in each round in a fashion that is conditional on the behavior of the group members
in previous rounds. This class of models is a natural extension to the group
structured population setting of tit-for-tat and related conditional 
strategies in the iterated two player setting.
We propose that this kind of conditional altruistic behavior in groups be investigated
as a possible route for the spread of altruistic behavior through natural 
selection. 

\end{abstract}


\newpage


\section{Core results}
\label{sec1}

We introduce a stylized framework for studying the evolution of a group 
structured population. Our goal is to shed light and clarify issues in the 
ongoing debate on the interplay between group selection and kin selection.
We will focus here on the application of the framework to the 
question of the spread of an altruistic gene A, resulting from a mutation, 
in the absence of further mutations. This is the central issue addressed 
in the debate, and is well suited for introducing our framework. Comments 
on other natural applications of the framework will be made 
at various places in this paper. 
For background material, and a significant sample of work addressing altruism,
cooperation, group selection and kin selection, from different perspectives, 
we refer the reader to the papers/books listed in the reference section 
(except for \cite{Dur} and \cite{Harris}) and references therein.

We conceived our approach in the spirit of basic stylized frameworks in 
population genetics, like the Fisher-Wright framework with selection. 
By this we mean that we
aimed at keeping the elements in the modeling mathematically precise, 
and as simple as possible, provided they would still capture the basic
biological features that one wants to study. Central to the contribution 
in the present
work is the fact that rigorous mathematical methods can be used 
to decide the fate of the mutant gene A. This allows us to compare our 
rigorous conditions for the spread of altruism with basic concepts and issues 
including Hamilton's rule, the Price equation, neighbor modulated fitness 
computations, the compatibility between a gene's eye view and a group selection 
mechanism, 
whether pairwise interactions, linearity of fitness functions, or 
weak selection have to be assumed,
the possibility of altruism to spread in group structured populations when the 
migration rate is significantly higher than the inverse of the group size,  
etc. We believe that 
our results help in 
clarifying these issues, and others that are being debated, and 
we hope that it will bring some consilience to this field, allowing for
a greater level of collaboration among the various groups contributing to
the area.

Our framework can be seen as a mathematically precise version of what in
\cite{LKWR}, p.6737, is
called a ``typical kin selection model''.
One of our main goals was to develop methods that apply to much more general fitness
functions than those considered there, and that do not require the assumption that
selection is weak.

Our framework is also a natural extension of the classical trait-group framework
(for the origins of this framework see \cite{Wil1})
and therefore allows for the analysis of the models that have previously been
studied in that framework. What distinguishes our framework from the 
trait-group one, is a migration rate parameter $0 \leq m \leq 1$, with the
case $m=1$ reducing to the trait-group framework. 
When $m < 1$, our framework introduces assortment, in the sense that
offsprings of members of the same group tend to stay together.
The migration parameter $m$ determines the strength of this assortment;
the smaller it is the stronger the assortment.

Under natural conditions on the altruistic gene A,
to be specified later, 
we show that there are two critical values of 
the migration rate $m$, namely $0 < m_f \leq m_s < 1$, playing roles as 
follows. For 
$m_s < m \leq 1$ the gene A is eliminated.
For $0 \leq m < m_f$ the gene A has a positive 
probability of fixating, replacing the wild allele N. 
In the intermediate regime, when $m_f < m < m_s$, 
the outcome is model dependent, but typically there is a 
positive probability for the mutant gene A to spread and reach 
a polymorphic equilibrium with the wild allele N. 
In this paper
we will focus on the critical point $m_s$ (the subscript $s$ 
stands for survival of the mutant allele A), and the corresponding 
mathematically rigorous conditions for the spread of altruism in our framework. 
Results on $m_f$ and the corresponding conditions for fixation of A, as well 
as results on the evolution of the frequency of the gene A in the population
when it spreads
(either in the intermediate polymorphic regime, or in the fixation regime)
will be presented elsewhere (\cite{SVC1}).

For the reader's benefit, and for brevity, we will not present here 
the full rigorous mathematical proofs, but rather explain why the various results
are true at a level that, we hope, will make them generally quite intuitive.
We emphasize that the results are mathematically 
rigorous and hold for any strength of selection. 
In the special case of weak selection simplifications occur and will also be 
discussed. 

\

We consider a population in which individuals live in a large number $g$ of groups of size $n$.
Individuals are of two genetically determined phenotypic types, the wild N and the altruistic 
mutant A. Reproduction is asexual and the type is inherited without mutation by the offsprings.
Each individual has a relative fitness that depends on its type and the types of the other members
of its group (the idea being that altruists, at a cost to themselves, provide a benefit to the 
members of their group). The relative fitness of an altruist, and that of a non-altruist, 
both in a group that has a total number $k$ of altruists, will be written, respectively, as
$$
w^A_k = 1 + \delta v^A_k,  \ \ \ \ \ \ \ w^N_k = 1 + \delta v^N_k,
$$
with the convention that $v^N_0 = 0$, i.e., $w^N_0 = 1$. 
The quantities $v^A_k$ and $v^N_k$ 
represent payoffs to altruistic, or non-altruistic behavior. The parameter $\delta \geq 0$ 
indicates the strength of selection, with the limit $\delta \to 0$ corresponding to the 
limit of weak selection, and the case $\delta = 0$ corresponding to the case in which there 
is no selection, only neutral genetic drift.
Examples of payoff functions will be provided in Section \ref{sec2}. See also Fig. \ref{fig2}. 

\begin{figure}[th!]
\begin{center}
\includegraphics[width=14cm]{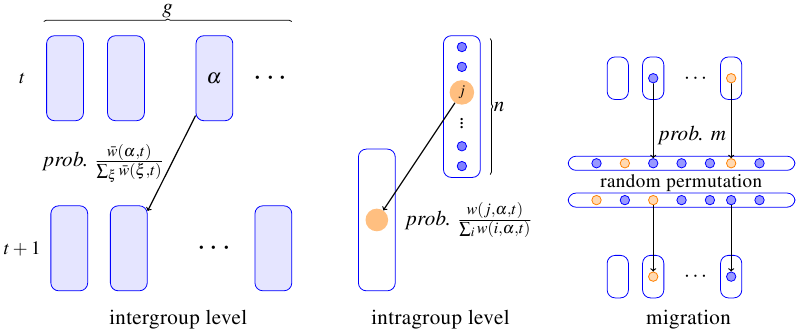}
\end{center}
\caption{Illustration summarizing the two-level Fisher-Wright process with selection and migration. Here $w(j,\alpha,t)$ is 
the fitness of individual $j$ of group $\alpha$ in generation $t$, and $\bar{w}(\alpha,t)$ is the average fitness of the 
members of group $\alpha$ in generation $t$. Intergroup level: in generation $t+1$ each and every one of $g$ groups choses a parent group 
$\alpha$ from the previous generation independently with probability proportional to $\bar{w}(\alpha,t)$. Intragroup level: 
each  individual inside a child group then independently choses an antecessor $j$ among $n$ individuals in his parent group with probability
 proportional to $w(j,\alpha,t)$ . Migration: each individual in each group is   marked as a migrant with probability $m$,
 migrants are then randomly shuffled.   }
\label{fig1}
\end{figure}

Evolution operates as
the next generation is formed through a process that involves group 
competition and competition within groups, followed by migration at rate 
$0 \leq m \leq 1$,
as summarized in Fig.\ref{fig1}. 
Competition among groups is idealized as an 
(intergroup-level) Fisher-Wright process with selection, described as follows. 
We associate to each group a relative fitness given by the average relative 
fitness of its members. This means that a group with $k$ altruists has
relative group fitness
\be
\bar w_k = \frac{k w^A_k + (n-k)w^N_k}{n} \ = \ 
1 + \delta \bar v_k,  \ \ \ \ \ \ \ \mbox{where} \ \ \ \bar v_k = \frac{k v^A_k + (n-k)v^N_k}{n}.
\label{barw}
\ee
Each group in the new generation 
has a parental group from the previous generation, chosen 
independently with probability 
proportional to group relative fitness. Competition 
among members of a group is described by an (intragroup-level) 
Fisher-Wright process with selection, described as follows.
The $n$ members of each group in the new generation each has a parent from among 
the $n$ individuals in their parental group, chosen independently 
with probability
proportional to the fitnesses of the members of that parental group.
(A standard probability computation, using conditioning on the parental
group, shows that each individual in the old generation has then an 
expected number of offspring proportional to its relative fitness. 
Conversely, for this to be true, the fitness associated to the groups in 
the intergroup level Fisher-Wright process must be given by (\ref{barw}). 
The conceptual relevance of this equivalence is emphasized in \cite{KG-S}.) 
Once the new $g$ groups have been formed according to this two-level 
Fisher-Wright process, a fraction $m$ of the individuals migrates from 
their group to a randomly chosen group, preserving the constancy of the 
number $n$ of members of the groups. More precisely, each individual, 
independently of anything else, leaves its group with probability $m$; 
the migrants then return to the $g$ groups in a random fashion, filling 
vacancies, so that each group has again $n$ members. 
Each possible way of assigning the migrants to 
the vacancies left in the groups is equally likely, meaning that the migration
process is completely random. (Mathematically: each individual is independently 
of anything else, with probability $m$, declared to be a migrant, 
and one applies then a random permutation to the set of migrants.) 

In the case $m=1$, we can equivalently think that the new groups are 
formed by random assortment from a metapopulation with $gn$ individuals. 
Each one of these $gn$ individuals has a parent chosen 
independently with probability
proportional to relative fitness from the $gn$ individuals in the 
old generation. This is precisely the traditional trait-group framework.

A model within our framework is specified by giving the values of $n$ and 
the relative fitnesses $w^A_k$ and $w^N_k$. The number of groups $g$ will
be considered to be very large, corresponding to taking the limit 
$g \to \infty$ in the computations. The study of how finiteness of $g$ 
modifies the conclusions is very interesting, but will be deferred 
to a later investigation. 


Our results on the viability of a single gene A to spread do not depend on any conditions
on the parameters of the model. (These results are summarized in the paragraph 
that contains display (\ref{ses}).) But to keep the presentation more focused and
interesting, we will assume that conditions (C1) and (C2) below hold, except when 
stated otherwise. 

\begin{description}

\item (C1) $v^A_1 < 0$, i.e., $w^A_1 < 1 = w^N_0$, so that an isolated
type A individual has lower fitness than the wild type N has in groups without
altruists.


\item (C2) $v^A_n > 0$, i.e., $w^A_n > 1 = w^N_0$, so that
type A individuals have greater fitness when in single-type groups than
type N individuals have when in single-type groups.

\end{description}
Condition (C1) is sometimes referred to, after \cite{Wil2}, 
as the condition for A to be 
called ``strongly altruistic''. This condition means that an isolated gene A
is at a disadvantage with respect to the wild type N in the population at large.
In the trait-group framework, $m = 1$, this condition makes it impossible 
for A to invade. We will see that, as expected, this condition 
makes it impossible for this gene to invade also when $m$ is close to 1, so that 
$m_s < 1$. 

On the other hand, we will see that condition (C2) is sufficient for A to 
spread when $m$ is close to 0, so that $m_s > 0$. 


\

We will study the evolution of the population, when started in generation 0
from the situation in which only one individual is of type A. Naturally 
this refers to the situation in which a mutation from N to A has just 
occurred, and to the assumption that the mutation rate  
is so low that no further mutations will occur before the fate of that 
mutant gene has been decided. Obviously the mutant A may disappear in a 
few generations, but we want to determine here when it is viable, in the 
sense that it has a fair chance 
of spreading. We will denote by $N^A(t)$ the number of altruists in 
generation $t$. 
In case $\delta = 0$, meaning that the mutation is neutral, 
the expected number of altruists remains constant, $\E(N^A(t)) = 1$, 
and a standard martingale argument gives probability $1/(ng)$ to the 
event that A will not disappear, but rather fixate eventually. As 
$g \to \infty$ this probability vanishes. When $\delta > 0$, and
condition (C1) holds, prospects are even worse for the mutant gene A in
generation 1. The expected number of type A individuals then is 
$\E(N^A_1) = w^A_1 < 1$. When $m=1$, these bad prospects worsen with time,
since in the first few generations the possible type A are likely to all be in 
different groups (since $g$ is large), and so are always carrying the same 
fitness $w^A_1 < 1$. This leads to $\E(N^A_t) = (w^A_1)^t$, and to the 
certain elimination of the gene A. In the opposite extreme, when $m = 0$, 
a group with $n$ altruists may be created by chance in a few generations. 
Groups that descend from this one will always have only type A individuals,
who therefore have average fitness $w^A_n > 1 = w^N_0$, by (C2). In this situation 
it is reasonable to expect that the altruistic gene can spread with a 
probability that does not vanish as $g \to \infty$. The rigorous analysis 
of what happens in this case and in the more important case $0 < m < 1$ 
can be done using the theory of multitype branching processes, as covered,
for instance in Chapter II of \cite{Harris}.
We turn next to the application of that theory to solving our problem. 

In applying multitype branching process theory here, we must emphasize that that theory
describes well the evolution in our framework only in its early 
stage. By early stage, to be abbreviated E.S., we mean the generations before the number of
groups that contain altruists is comparable to $g$. We will nevertheless
see that this early stage period is of order $\log g$ generations,
so that it covers a large number of generations since $g$ is large. 

We say that a group is of type $k$ if it has exactly $k$ altruists. 
First we explain how multitype branching theory can be used when $m = 0$.
In the E.S., there are few groups with altruists, compared with 
the total number $g$ of groups. Therefore in the intergroup Fisher-Wright
process the competition among groups with altruists is basically irrelevant;
groups are mostly competing with the groups without altruists, that form 
the background on which groups with altruists may or not spread. To see
this, note first that in each generation 
there are $g$ new groups being formed, and that they
choose their parental groups independently with probability proportional to
group fitness. Since the vast majority of the groups have no altruists, 
and therefore group fitness $\bar w_0 = w^N_0 = 1$, a type $k$ group has a
probability close to $\bar w_k / g$ of being the parent of each new group. 
Hence, each 
group of type $k$, with $k \geq 1$, can be seen, in first approximation, 
as creating independently of the other ones a number of offspring groups
that is given by a binomial distribution with parameters $g$ and 
$\bar w_k / g$ (well approximated by a Poisson distribution with mean 
$\bar w_k$). 
During the E.S.,
there are much less than $g$ groups with altruists, 
and they are each producing a number of 
offspring groups that is also small compared to $g$. For this reason 
each one of these groups interferes little with the other groups with 
altruists in their creation of offspring groups. This independence in
the creation of offspring groups is what defines a multitype branching 
process. 
The next fact to observe is that each group that has as its 
parental group a group of type $k$ will be of type $k'$, due to the 
intragroup Fisher-Wright process, with probability
$$
p(k,k') = \P \, (\mbox{Bin}(n,k w^A_k / n \bar w_k) = k'),
$$
where $\mbox{Bin}(n,p)$ is a binomial random variable with $n$ 
attempts, each with probability $p$ of success. 
Assembling the pieces above, we conclude that, through the two-level Fisher-Wright  
process, a group of type $k$ creates in the average 
\be
M_{k,k'} = \bar w_k \, p(k,k') 
\label{M}
\ee
groups of type $k'$ in the next generation, independently of anything else. 
When $m = 0$ this is the whole story. The matrix $M$, of size $n \times n$,
defined by (\ref{M}), with $k = 1, ..., n$, $k' = 1,...,n$, 
characterizes the evolution of this process 

When $m > 0$, the creation of the new generation of groups is complicated
by migration. One could be concerned that a multitype branching process 
description is no longer feasible. Fortunately this fear is unfounded,
thanks to the fact that we are only considering the E.S., during 
which $N^A(t) << g$. Altruists form then a minute fraction of 
the migrant population, and as a consequence it is unlikely that migrant
altruists will settle in groups that contributed altruists to the migrant
population, or that any group will receive more than one migrant altruist.
A group that has $k'$ altruists before migration, will keep after migration
a random number of altruists given by a binomial distribution with $k'$ 
attempts and probability $1-m$ of success. 
This means that the probability that after migration this group is replaced
by a group with $k''$ altruists is given by
\be
A_{k',k''} = \P \, (\mbox{Bin}(k',1-m) = k''). 
\label{A}
\ee
That group that had $k'$ altruists before migration, will also
be contributing with
an expected number $k' m$ of migrant altruists, who are likely each 
to settle in a different group that had no altruists before migration, and has 
exactly one altruist after migration, i.e., is now of type 1. 
This means that the expected number of groups of type $k''$ created from groups of
type 0 that received altruists from our group that had $k'$ altruists before 
migration is given by
\be
B_{k',k''} \ = \
\left\{
\begin{array}{cc}
                    m k', & \mbox{if \  $k'' = 1$,} \\
                    0,    & \mbox{otherwise.}
\end{array}
\right.
\ee
The matrix $M$ should therefore be replaced, due to migration,
with the matrix $M(A+B)$ in 
describing the expected number of groups of type $k''$ created 
in the new generation by each group of type $k$ in the old generation. 
We will use the notation $N_k(t)$ for the 
number of groups of type $k$ in generation $t$, and also write  
$N(t) = (N_1(t), ..., N_n(t))$. In summary, we have, in matrix notation, 
that for $t$ in the E.S., 
\be 
\E \, (N(t+1) \, | \, N(t)) \ = \ N(t) \, M \, (A+B).
\label{NM} 
\ee

Obviously $N^A(t) = N_1(t) + 2N_2(t) + ... +n N_n(t)$. Therefore,
the survival of the altruistic gene is equivalent to the survival of the 
multitype branching process $N(t)$.
Next we describe the necessary and sufficient condition for the survival with positive
probability of this multitype branching process.
In what follows we will suppose that $0 < m < 1$; the cases $m=0$ and
and $m=1$ can be treated as limits.
Because the matrix $M(A+B)$ has then only strictly positive entries, 
it results from the Perron-Frobenius Theorem 
(see, e.g., Theorem 5.1 in Chapter II of \cite{Harris})
that it has an eigenvalue 
$\rho$ that is simple, positive and larger in absolute value than all other eigenvalues.
It corresponds to left and right eigenvectors, both
of which have all their entries strictly positive. We will 
denote by $\nu$ this left-eigenvector, normalized 
so as to represent a probability distribution over group types: $\nu_1 + ... + \nu_n = 1$. 
(Illustrations of $\rho$ and $\nu$ as functions of $m$ and $\delta$ appear 
in Fig. \ref{fig3}, Fig. \ref{fig5} and Fig. \ref{fig8}, for various models.)
A consequence of (\ref{NM}) and of 
the Perron-Frobenius Theorem
is that for $t$ large, but still in the E.S., 
\be
\E N(t)  \ =  \ C \rho^t \nu,  
\ \ \ \ \ \ \ \ 
\E N^A(t)  \  =  \ C \rho^t \sum_k k \nu_k.
\label{ENrhot}
\ee
where $C>0$ is a constant.

Theorem 7.1 
in Chapter II of \cite{Harris}
states that the survival with positive probability
of the multitype branching process is equivalent to the condition 
\be
\rho(m) > 1.
\label{rho>1}
\ee
(We will
make $m$, $\delta$, etc, explicit in the notation $\rho$,
$\nu$, etc, only when important.) 
The critical value $m_s$ is then obtained by solving the equation
\be
\rho(m_s) = 1. 
\label{rhoms}
\ee
(For some numerical examples, see Fig. \ref{fig3}, Fig. \ref{fig4}, Fig. \ref{fig6} and Fig. \ref{fig7}.)
To see that this equation has a solution in the open interval $(0,1)$, 
we use continuity of eigenvalues and eigenvectors. In particular $\rho(m)$ is 
continuous, and it is enough to observe how it behaves as $m$ approaches 
$0$ or $1$.
When $m = 0$, $M(A+B) = M$ has the eigenvalue $w^A_n > 1$ (by (C2)),
corresponding to the left-eigenvector $(0,0,...,0,1)$. This implies that $\rho(m)$ 
must be larger than 1 when $m$ is close to 0. When $m = 1$, the matrix $A=0$, and 
$M(A+ B) = MB$ has only the first column not identically 0. Therefore any of its 
left-eigenvectors must be of the form $(a,0,...,0)$. When such a vector is multiplied 
by $MB$, the result is $w^A_1 (a,0,...,0)$. This shows that $w^A_1$ is the 
only eigenvalue of $M(A+B)$. Therefore $\rho(m)$ must converge to $w^A_1 < 1$
(by (C1)), as $m \to 1$. 

The argument above shows that (\ref{rhoms}) has a solution. Uniqueness of this solution 
is not guaranteed, unless additional conditions are assumed on the fitnesses. 
In any case, if there 
is more than one solution to (\ref{rhoms}), the natural definition of $m_s$, that we
adopt, is as the largest one, representing the least strength of assortment of altruists that 
suffices to allow altruism to survive.

The theory of multitype branching processes provides us with further detailed 
information on the patterns of evolution, when the altruistic gene A survives. 
Theorem 9.2 of Chapter II of \cite{Harris} shows that in the event that
the process survives, it behaves in a rather regular fashion: as $t$  
becomes large, the vector $N(t)$ tends to become a multiple of $\nu$, and to grow
at rate $\rho$. (More precisely, the distribution of the random vector
$N(t)/\rho^t$ converges to $Z \nu$, where $Z$ is a random variable.)
Intuitively, this is a sort of law of large numbers:
if the multitype branching process $N(t)$ survives, the 
relative frequency of groups of each type tends to stabilize, as that
given by the vector $\nu$, as $N^A(t)$ becomes large.
But randomness in the values of $N_1(t)$,...,
$N_n(t)$ persist (and are given by the one dimensional $Z$ above), due to the randomness 
that affects the process in the first few generations, before large 
number phenomena can take place.

We summarize now what we have learned about the evolution of the gene A in
our framework. When (\ref{rho>1}) fails this gene dies out in a few
generations. On the other hand, when (\ref{rho>1}) holds, the picture of
its evolution is a dichotomy. Either A is eliminated in a few generations, 
or else it survives and, as it spreads, its distribution  
stabilizes dynamically, in the sense that
\be
 N(t) \ = \  \rho^t Z \nu,
\ \ \ \ \ \ \ \
N^A(t) \ = \ \rho^t Z \sum_{k \geq 1} k \nu_k,
\label{ses}
\ee
when $1 << t << \log g / \log \rho$. We will refer to this time period as 
the stationary early stage, abbreviated S.E.S..
Note that the upper bound on the magnitude of
$t$ is equivalent to the condition that during this period $N^A(t) << g$.
The condition $ t << \log g / \log \rho$ is then what defines the E.S., and 
we will refer to the first few generations, before the S.E.S., as 
the very early stage, abbreviated V.E.S.. 
During the V.E.S., $N(t)$ evolves very randomly, displaying little 
regularity, since the number of copies of the gene A is small. 
The random variable $Z$ reflects how randomness during that period 
affects the later S.E.S., and the extent to which it is not washed out 
as the number of copies of A grows and the evolution becomes more regular.

Finally, the later period when 
$N^A(t)$ is no longer negligible as compared to $g$ (provided that A has survived) 
will be called late stage, abbreviated L.S.. The evolution of $N(t)$ during 
the L.S. will no longer be well approximated by the multitype branching 
process, and is more challenging to study. To focus in the current paper
only on the issue of survival of the gene A, we will postpone our analysis 
of that problem to a later publication (\cite{SVC1}). Here we only observe that in that 
regime, laws of large numbers allows us to well describe the evolution as a 
dynamical system (in $n$ dimensions, representing the fractions of groups
of each type). That dynamical system turns out to be non-linear (due to 
migration) and to sometimes have more than one stable equilibrium. Nevertheless,
under the condition that (C2) holds and $\delta>0$ is small, or some 
alternative conditions (for instance (C3) in the next section),
when $m>0$ is small, it has a single stable equilibrium,
corresponding to fixation of the gene A in the population. This is what characterizes
the critical point $m_f$. It is worthwhile to stress in this connection 
that the linearity of the evolution during the E.S., as given by (\ref{NM}), 
in spite of migration, makes the issue of deciding when the mutant A can survive much easier 
than it would be otherwise. This is what allows the reduction of the problem 
to a standard eigenvalue problem.  


Returning to our analysis of the E.S., 
We can see (\ref{ses}) as a form of self-organization of the gene A.
If it survives, it arranges itself according to
the distribution $\nu$, that is a left-eigenvector of the driving matrix
$M(A+B)$. Left-eigenvectors are precisely the arrangements that the process
can have which are preserved in time. To better appreciate what is special about $\nu$,
and for several future uses,
we observe that if $\nu'$ is a left-eigenvector of $M(A+B)$, with eigenvalue
$\rho'$, and no negative entries, then
\be
\rho' \ = \ \frac{\sum_k \, w^A_k \, k \, \nu'_k}{\sum_k \, k \, \nu'_k}.
\label{rhonu'}
\ee
In other words, $\rho'$ is the average fitness of the individuals who carry
the gene A (or simply, the average fitness of the gene A), when the groups
with altruists are distributed according to $\nu'$.
Identity (\ref{rhonu'}) is an easy consequence of two observations.
First that from (\ref{NM}) we know that if $N(t) = C \nu'$ for some
constant $C$,
then $\E N(t+1) = C \rho' \nu'$. This implies that
$\E N^A(t+1) = C \rho' \sum_k k \nu'_k$.
Second, that in the multitype branching process an individual who carries
the gene A and belongs to a group with $k$ altruists produces an expected
number of offspring $w^A_k$, so that we can also write
$\E N^A(t+1) = C \sum_k w^A_k k \nu'_k$.
Comparison of these expressions yields (\ref{rhonu'}).

We combine (\ref{rho>1}) with (\ref{rhonu'}) to write the necessary and
sufficient condition for viability of the gene A as
\be
\frac{\sum_k \, w^A_k \, k \, \nu_k}{\sum_k \, k \, \nu_k} \ =
\rho \ > \ 1.
\label{rho>1nu}
\ee

Since $\nu$ is the only left-eigenvector of the maximal eigenvalue $\rho$,
(\ref{ses}) is telling us that when the altruistic gene survives, it tends
to organize itself 
(or we can also say ``nature organizes it, through natural selection'')
in the stable way that maximizes its average fitness.
This observation also makes the condition of survival (\ref{rho>1nu}) look
particularly natural. If $\rho \leq 1$, there is no stable way for the
gene A to be organized so that it will have an average fitness that is
larger than that of the wild type N in the population at large, where A
is still rare; A will then not be viable. On the other hand, when
$\rho > 1$, the gene A can be organized according to $\nu$, that is
stable, and provides it with mean fitness larger than 1, as needed for it
to spread among the wild type N.

It is enlightening to see what happens when the inequality in 
(\ref{rho>1nu}) fails. Even in 
this case, chance may produce at a time $t$ during the V.E.S. the arrangement 
$N(t)=(0,0,...,0,1)$, meaning that there are exactly $n$ altruists, all 
in the same group. Condition (C2) tells us that at time $t$ the average
fitness of the gene A is larger than 1. And indeed, 
$\E N^A(t+1) = w^A_n n > n = N^A(t)$, so that the altruistic gene is spreading
at this time. But this arrangement is not stable.  
In successive generations, the distribution of 
$N(t)$ is driven to a combination of left-eigenvalues of $M(A+B)$, and
$N^A(t)$ can grow then at most at rate $\rho \leq 1$, so that eventually
it is eliminated. 

In contrast, when (\ref{rho>1nu}) holds, chance will dictate if during the 
V.E.S. an arrangement of copies of A will form that not only 
provides that gene with mean fitness larger than 1, but is also likely to 
produce a succession of arrangements in the next generations, all with 
this property. The important point is that, under (\ref{rho>1nu}),
such arrangements do exist, and drive the evolution towards $\nu$.

The contrast in the last two paragraphs is one of the main lessons from 
our analysis. This lesson goes beyond the specific aspects of the stylization
that we are adopting here. We suggest making this idea explicit 
as a guiding criterion, that should be of use when considering
any framework, model or experimental situation. 

\

\noindent{\it Viability or Survivability Criterion:}
In a large population, 
a single mutant gene A, in the absence of further mutations,
will be viable, i.e., will have a positive probability of 
surviving and spreading, if and only if
this gene can produce in a few generations 
an arrangement of its copies in a number that is 
still small compared to the size of the population,  
but is likely to produce in the next generations a
sequence of arrangements with a growing number of its copies, until 
it accounts for a non-negligible fraction of the alleles in the population. 

\

We will refer to such a sequence of arrangements as a {\it survival mechanism}
for the mutant gene A, 
so that the criterion stated above postulates the existence of a survival mechanism
as a necessary and sufficient condition for the viability of the mutant gene A.
Such a mechanism can, for instance, be started 
by an arrangement of copies of the gene A
that satisfied the three conditions below:

\begin{description}

\item(i) When in this arrangement the average fitness of the mutant gene is larger
than that of the wild type in the population at large, before the mutant appeared.

\item(ii) This arrangement is likely to produce in the next generation another
arrangement with the same property (i) above.

\item(iii) Due to the growth in the number of copies of the gene, the probability 
of success in step (ii) above increases from generation to generation,
fast enough to assure that 
the probability of producing the sequence of arrangements
mentioned in the criterion is large.

\end{description}
Indeed, 
in our framework, once an arrangement of copies of A is produced with distribution in 
groups close to $\nu$, conditions (i), (ii), (iii) are fulfilled, provided that 
$\rho > 1$. On the other hand, when $\rho \leq 1$, no arrangement exists that satisfies
these three conditions.

\

We end this section with some observations on generalizations of our methods. One can 
modify the intergroup and the intragroup competition procedures, from the Fisher-Wright 
ones that we consider here, and in this way extend our framework further. 
For instance, modifications to the intragroup selection procedure can be fairly 
general, and would only require a modification of the matrix $p(k,k')$. Instances
of such a modification could include domination patterns in the intragroup reproduction
mechanism, that result in reproductive skew within the group. In an extreme case, 
a single member of the group, chosen at random with probabilities proportional to 
individual fitness of the group members, could mother all the $n$ offspring of an offspring 
group. 

Modifications of the intergroup competition mechanism are even simpler to consider. Note 
that we did not use in our analysis the full power of the assumption that this 
mechanism is a Fisher-Wright procedure. We only assumed that if altruists are rare, 
then groups with altruists father each in the next generation an almost independent
random number of groups, with mean proportional to group fitness (defined as average 
fitness of group members). Under these broad assumptions our methods and results above,
and in the remainder of this paper, are unchanged. We chose to introduce our framework
with a Fisher-Wright competition mechanism among groups for concreteness. 
This choice forces the number of offspring groups of each group to be binomially distributed 
(well approximated by Poisson, since $g$ is large). 
The observation in the current paragraph is of special relevance then 
in situations in which the number 
of groups fathered by each group is better modeled by a distribution that  
is far from Poisson, as for instance in cases in which their variance is 
much smaller than their mean 
(as happens, for example, when $\delta$ is small, the mean is close 
to 1, and the variance is much smaller than 1, with most groups fathering 
exactly one group). 
 
\section{Models}
\label{sec2}

\begin{figure}[th!]
\begin{center}
\includegraphics[width=12cm]{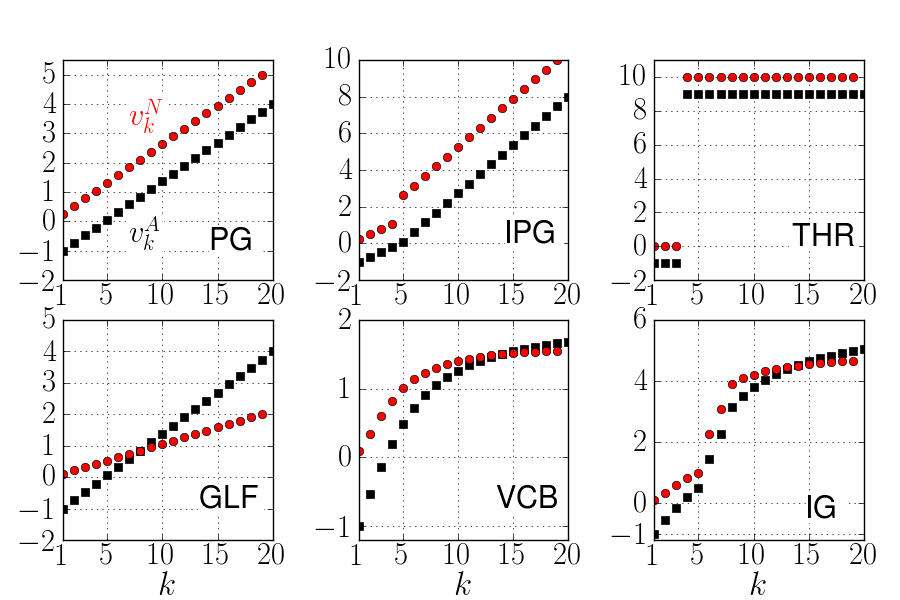}
\end{center}
\caption{Payoff profiles. Payoffs for the wild  (``non-altruist'', N) type $v^N_k$ are represented as red 
circles while black squares 
depict payoffs $v^A_k$ for the mutant (``altruist'', A) type.  From top left: Public goods game (PG, Example 1) for $n=20$, 
$C=1$ and $B=5$. Iterated public goods game (IPG, Example 2) for $n=20$, $C=1$, $B=5$, $a=4$ and $T=10$. 
Threshold model (THR, Example 3) for $n=20$ $C=1$, $\theta=4$ and $A=A^\prime=10$. General linear fitness 
(GLF, Example 4) for $n=20$, $C=1$, $B=5$ and $B^\prime=2$. Variable costs and benefits (VCB, Example 5) 
with $C_k=C/k^{a_1}$, $B_k=bk^2/(1+dk^2)$ and $B^\prime_k=b^\prime k^2/(1+d^\prime k^2)$ for $n=20$, $C=1$, $a_1=0.5$, $b=b^\prime=2$, $d=0.05$, $d^\prime=0.065$. 
Iterated game (IG, Example 6) with cost and benefit functions as in the VCB case and $T_k=1$ if $k\le 5$, $T_6=2$, $T_7=2.5$
 and $T_k=3$ if $k\ge8$. }
\label{fig2}
\end{figure}

\begin{figure}[th!]
\begin{center}
\includegraphics[width=12cm]{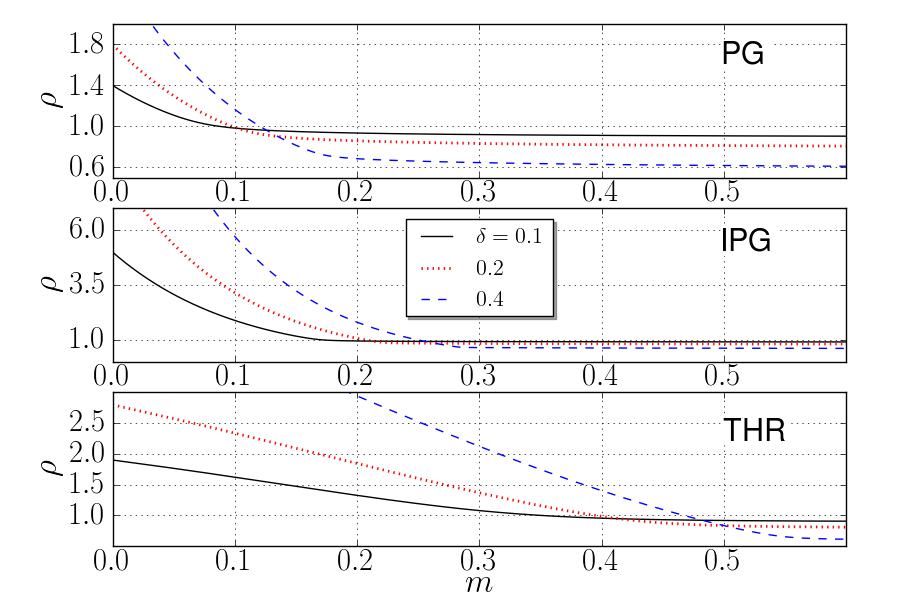}
\end{center}
\caption{Perron-Frobenius eigenvalues $\rho$ as a function of $m$ for $\delta=0.1$, $0.2$ and $0.4$. From top to 
bottom: Public goods game (PG, Example 1) with $n=20$, $C=1$, $B=5$. Iterated public goods 
(IPG, Example 2) with $n=20$, $C=1$, $B=5$, $a=8$ and $T=10$. Threshold model (THR, Example 3) 
with $n=20$, $C=1$, $\theta=4$, $A=A^\prime=10$. Critical migration values $m_s$ are obtained by solving 
$\rho(m_s)=1$. These figures should be, respectively, compared to Figure \ref{fig4}, Panel  B, blue dashed line (PG);
 Figure \ref{fig5}, Panel B, blue dashed  line (IPG); and Figure \ref{fig6}, Panel A, blue dashed line (THR). 
}
\label{fig3}
\end{figure}

\begin{figure}[th!]
\begin{center}
\includegraphics[width=12cm]{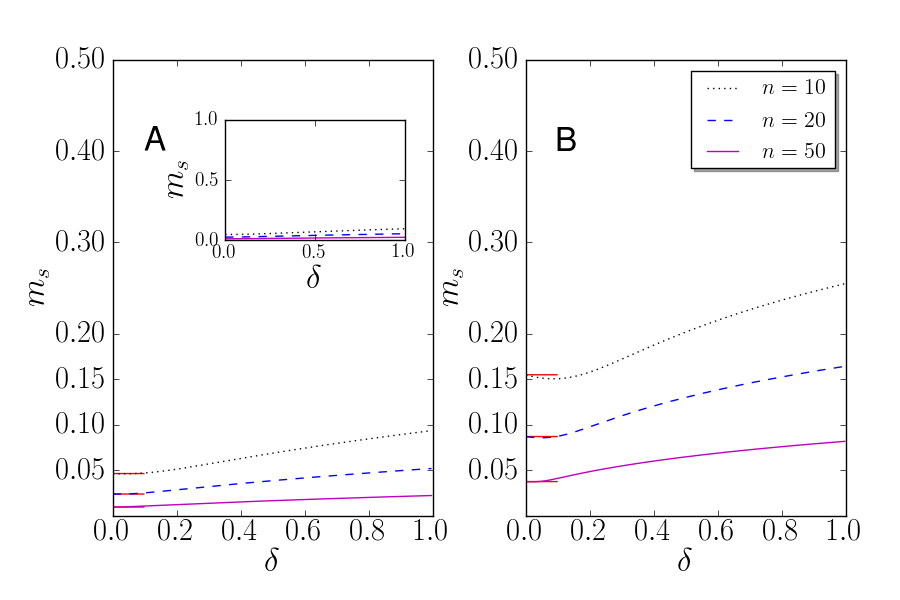}
\end{center}
\caption{Public goods game (Example 1): Panel A represents critical values $m_s$ as a function of the strength of 
selection $\delta$. Curves correspond to the case $C=1$, $B=2$ and $n=10$ (top, black dotted line), $n=20$ 
(middle, blue dashed line) and $n=50$ (bottom, magenta full line). Red lines indicate critical values at the weak selection 
limit obtained from the viability condition (\ref{HR+wspi}), or (\ref{R0sforEX1}). The inset shows the same curves within the full range of 
possible values for $m_s$. Panel B depicts the same conditions except for $B=5$.}
\label{fig4}
\end{figure}

\begin{figure}[th!]
\begin{center}
\includegraphics[width=12cm]{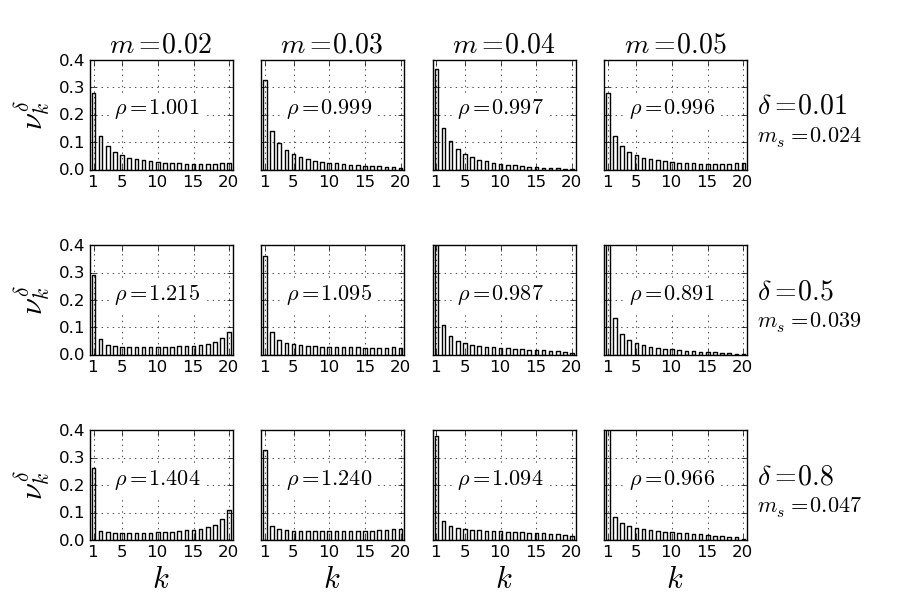}
\end{center}
\caption{Public goods game (Example 1): Perron-Frobenius eigenvectors $\nu_k$ as a function of the strength of selection
$\delta$ (rows) and of the migration rate parameter $m$ (columns). Critical migration rates $m_s^\delta$ 
are annotated in each row. Perron-Frobenius eigenvalues $\rho^\delta(m)$ are also provided for each case.
  Histograms represent the case $C=1$, $B=2$ and $n=20$.}
\label{fig5}
\end{figure}

\begin{figure}[th!]
\begin{center}
\includegraphics[width=12cm]{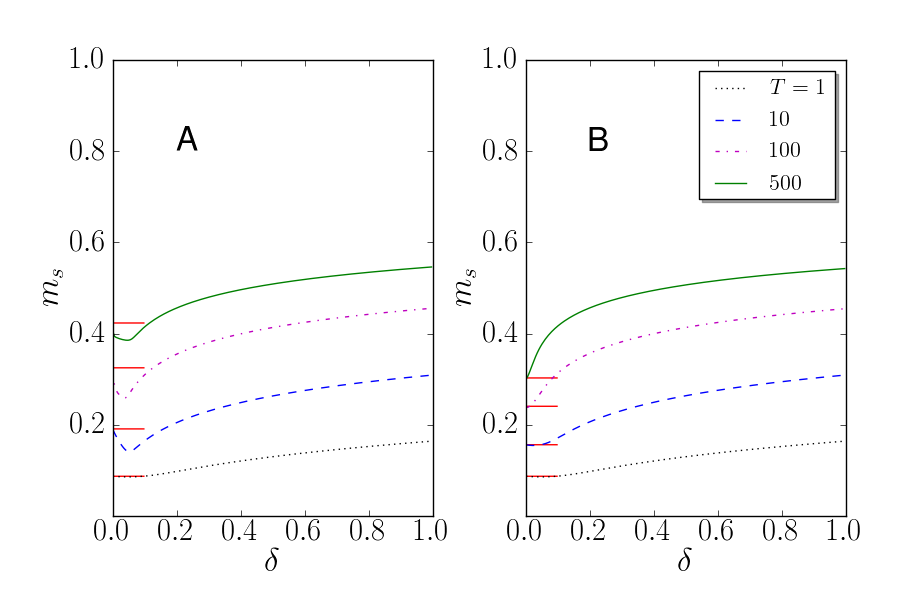}
\end{center}
\caption{Iterated public goods game (Example 2): Critical values $m_s$ as a function of the strength of selection $\delta$. 
Panel A depicts the case $n=20$, $C=1$, $B=5$, $a=4$ with, respectively from bottom to top, $T=1$ (dotted black line), $T=10$ (dashed blue line),
 $T=100$ (dot-dashed magenta) and $T=500$ (green full line). Panel B depicts the same conditions except for $a=8$.
Red lines indicate critical values at the weak selection limit obtained from the viability condition (\ref{HR+wspi}).}
\label{fig6}
\end{figure}

\begin{figure}[th!]
\begin{center}
\includegraphics[width=12cm]{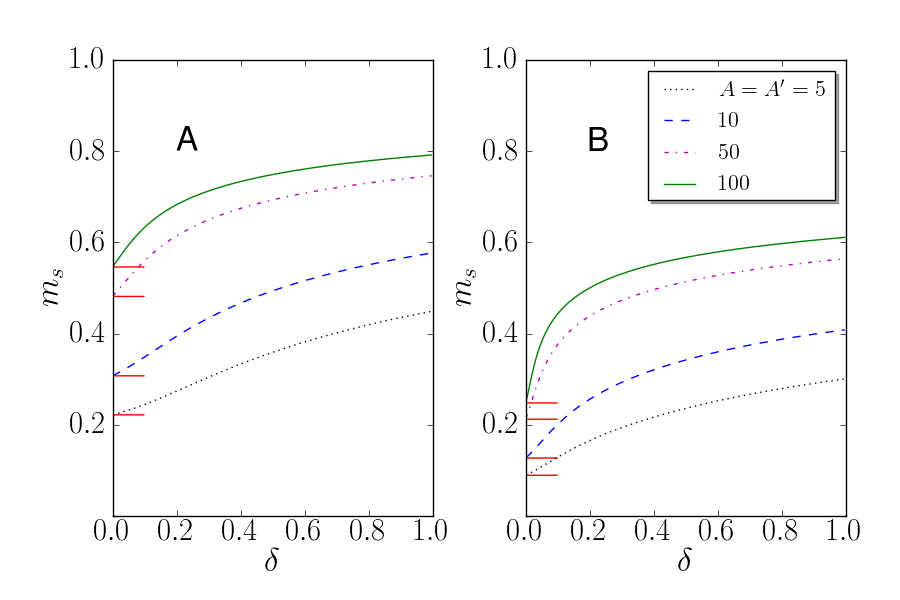}
\end{center}
\caption{Threshold model (Example 3): Critical values $m_s$ as a function of the strength of selection $\delta$. 
Panel A depicts the case $n=20$, $C=1$, $\theta=4$ with, respectively from bottom to top, $A=A^\prime=5$ (dotted black line), 
$A=A^\prime=10$ (dashed blue line),
 $A=A^\prime=50$ (dot-dashed magenta) and $A=A^\prime=100$ (green full line). Panel B depicts the same conditions except
 for $\theta=8$. Red lines indicate critical values at the weak selection limit obtained from the viability condition 
(\ref{HR+wspi}), or (\ref{HRtmws}).}
\label{fig7}
\end{figure}

In this section we will introduce several models and discuss their relevance. 
Fig. \ref{fig2} provides an overview of some of their typical features. Fig. \ref{fig4}, Fig. \ref{fig6} 
and Fig. \ref{fig7} provide values of $m_s$ as function of the strength of selection 
$\delta$ for some of them. Notice, from these figures, that $m_s$ is not always
monotone in $\delta$, but that it often increases substantially when $\delta$
is large. This fact highlights the relevance of studying the models not only 
when selection is weak. Notice also that in Fig. \ref{fig6} and Fig. \ref{fig7}, the product $n m_s$
can be of the order of $10$. This is relevant in view of the widespread claim that 
altruism cannot survive when $n m$ is significantly larger than $1$. As far as we 
know, this perception resulted from the analysis of particular models (e.g., 
in \cite{MS2}, \cite{Aoki} and \cite{CA}) and an excessive emphasis on the 
public goods game (Example 1, below). 
One of the main messages from this section and the following ones 
will, indeed, be that mechanisms 
that go beyond the public goods game may be central to the understanding of the
spread of altruistic genes, and can be analyzed in our framework 
with no special difficulty.
 
\

Conditions (C1) and (C2) are very mild, and basically characterize the effect
of the gene A on its carriers as an individually beneficial social effect which
comes with a personal cost to them.
One does not have to restrict oneself to behavioral effects of the
gene A on its carriers phenotype.
For example, another kind of application could include
anatomic and physiologic effects that carry a cost, but produce benefits to those with
the altered phenotype when
in groups with others that share this feature. For instance, gene A could promote changes that
facilitate verbal communication with others that have the same changes, but at a cost,
say, in adding expensive tissue to the brain. An isolated carrier of gene A would
suffer the costs of carrying it, but without the possibility of benefiting from its
potential advantages.

In order to assure $m_f > 0$, in a forthcoming paper (\cite{SVC1}),
we will either have to assume that in addition to (C2) holding, $\delta$ is small, or else 
we will need to add an additional assumption. A sufficient one will be:
\begin{description}
\item (C3) $v^A_n = \bar v_n \geq \bar v_k$, i.e.,
$w^A_n = \bar w_n \geq \bar w_k$, for $k=0, 1,...,n-1$, so that the average fitness 
of a group is maximized when the group contains only altruists.
\end{description}

When the gene A affects behavior, what
precise conditions on the fitnesses $w^A_k$ and $w^N_k$ should be required for 
this gene to be called ``altruistic''? There is no agreement on the answer. 
The issues are very nicely presented and discussed in \cite{KG-SF}. We list next a few
of the conditions that can naturally be associated with altruism. These ones, and
a few more can be found in \cite{KG-SF}, where detailed references and credit are given.

Some of the conditions require 
the altruistic behavior to be beneficial.
Typical conditions of this kind are:

\begin{description}
\item (C4) $v^A_k$, or equivalently, $w^A_k$, is increasing in $k = 1,...,n$,
so that altruists are always better off sharing their group with more altruists.

\item (C5) $v^N_k$, or equivalently, $w^N_k$, is increasing in $k=0,...,n-1$,
so that non-altruists are always better off sharing their group with more altruists.

\item (C6) $\bar v_k$, or equivalently, $\bar w_k$, is increasing in $k=0,...,n$,
so that the members of a group are in the average better off with more altruists in
the group.
(Note that (C6) is an extension of (C3).) 

\end{description}

And complementary conditions require 
the altruistic behavior to come at a cost to the actor:

\begin{description}
\item (C7) $v^A_k < v^N_k$, i.e., $w^A_k < w^N_k$, $k=1,...,n-1$,
so that altruists are always worse off than non-altruists in the same group.

\item (C8) $v^A_{k+1} < v^N_k$, i.e., $w^A_{k+1} < w^N_k$, $k=0,...,n-1$,
so that an individual that suffered a mutation from N to A, would be worse off.
\end{description}
Condition (C8) extends condition (C1). It is known 
that when it holds,
then in the trait-group framework, $m=1$, starting from any fraction $p < 1$ of
genes A in the population, these genes will be eliminated.
(\cite{KG-SF} attributes this result to \cite{MJ}.)
The assortment provided by a low $m$ is nevertheless sufficient to allow
a single altruistic gene A to invade, if condition (C2) holds.


\

We will illustrate the use of our framework with several examples (see Fig.\ref{fig2}):


\

\noindent  {\bf Example 1.} {\it Public goods game:}

\begin{eqnarray*}
v^A_k & = & -C + (k-1)B/(n-1),  \\
v^N_k & = &    k B/(n-1),
\end{eqnarray*}
for positive constants $C$ and $B$. 
One can think that at a cost $C$ to itself, each altruist 
provides a benefit $B/(n-1)$ to each one of the other members of its group. 
Alternatively, one can think that at a cost $c$ to itself, each altruist
provides a benefit $b/n$ to each member of its group, itself included. Set 
then $C = c -b/n$, for the net cost to the altruist, and $B = b(n-1)/n$, 
for the total benefit to the other members of the group.
Each one of these two descriptions is common in nature,  
has its theoretical advantages and both appear often in the literature (see \cite{Pepper}
for more on this point).
The former description is often referred to as ``other-only'' trait, and the
latter one is then referred to as ``whole-group'' trait. Their mathematical 
equivalence illustrates something that presents itself a number of times. Two models
may be different in relevant biologic aspects, 
but lead to the same functions $v^A_k$ and $v^N_k$,
possibly after some change of variables, as above. In this case, we will say that
the models are materially different, but formally equivalent. 

There is a second way in which the present example splits into two materially
different, but formally equivalent descriptions. On one hand,
altruists could be performing individual actions, producing 
identical benefits to all the other (or to all, self included) members of the group. 
In some applications this may 
be a good description of what is happening. For instance, altruists could be individuals
with a hygiene habit that is beneficial to the group, in preventing disease, but 
costly to the actor. Alarm calls are another example. 

On the other hand,
the fitness functions in this example can also accommodate collective actions, 
in which altruists act together, to produce a common good for the group. Fighting in
a war against another group, or participating in collective hunting activities (with the 
product of the hunt shared among all members of the group) would
be examples of this kind. 

In all cases, the assumptions that the total benefit 
produced, $Bk$, grows linearly with the number $k$ of actors, and the net cost to each actor, $C$, is 
constant, may be unrealistic. We will explore these points in Example 5, below.

Note that $\bar v_k = k(B-C)/n$.
Condition (C1) holds since $v^A_1 = -C < 0$.
We suppose that $C < B$, so that (C2) holds, since $v^A_n = B-C$. Note that 
then also (C3)-(C8) all hold. 

\begin{figure}[th!]
\begin{center}
\includegraphics[width=12cm]{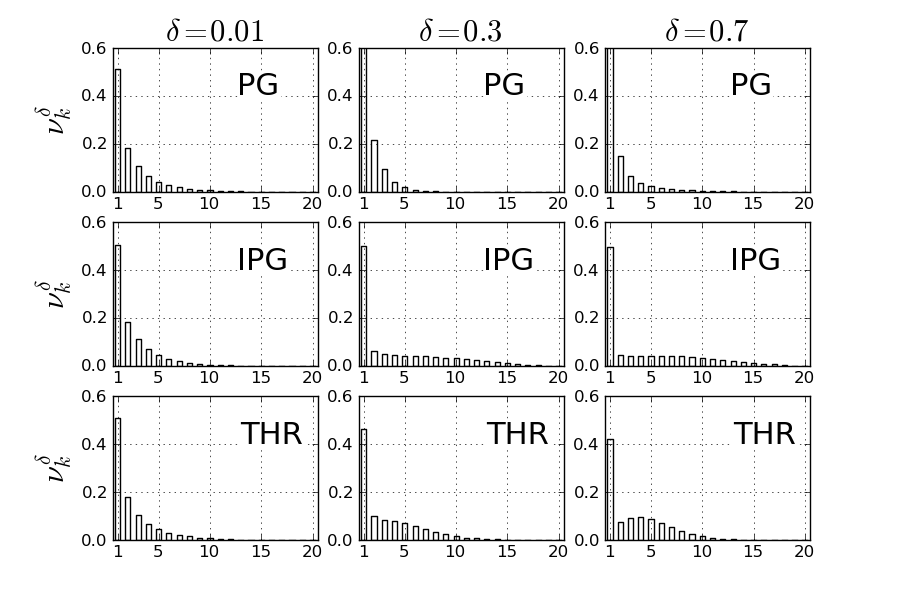}
\end{center}
\caption{Perron-Frobenius eigenvectors $\nu_k$ for selection strengths $\delta=0.01$ (left column), $\delta=0.3$ (middle column) 
 and $\delta=0.7$ (right column). Migration rate is set to $m=0.1$ and group sizes to $n=20$. Each line represents a different model. The top row, labeled as 
PG depicts the Public Goods game (Example 1) with parameters $C=1$ and $B=2$. The Iterated Public Goods game (Example 2) with parameters 
$C=1$, $B=4$, $a=4$ and $T=10$ is shown in row at the middle, labeled as IPG. The bottom row shows Perron-Frobenius
 eigenvectors for the Threshold model (THR, Example 3) with $C=1$, $A=A^\prime=5$ and $\theta=4$.  
The leftmost column emphasizes that the weak selection limit  $\nu_k^0$ is  independent of the model. In contrast, 
when selection is strong, $\nu^\delta_k$ depends on the model, as illustrated in the other columns.}
\label{fig8}
\end{figure}

Numerical results for Example 1 appear in Fig. \ref{fig4}, Fig. \ref{fig5} and Fig. \ref{fig8}. 

The public goods game has rightfully been called ``the mother of all 
cooperative models'' (see footnote 1 in \cite{Vee1}).
It is natural to study its behavior, and to understand how the gene A
can spread in this case.  In the next examples we will nevertheless 
try to convey the message that one should aim at 
developing methods, as we do here,
that can address more general models, as well. 
In the spirit of that metaphor, it is natural to see the next example as 
``a special daughter of the public goods game. It derives from the public goods game
in the same way that (in a two player setting)
the iterated prisoner dilemma and the tit-for-tat strategy
derive from a one shot prisoner dilemma game and a simple cooperative strategy.


\

\noindent  {\bf Example 2.} {\it Iterated public goods game. Altruists cooperate conditionally:}

\begin{eqnarray*}
v^A_k & = & \left\{
                \begin{array}{cc}
                                    -C + (k-1)B/(n-1),  & \mbox{if \  $k \leq a$},  \\
                                   T \, (-C + (k-1)B/(n-1)), & \mbox{if \  $k > a$}, 
                \end{array}
\right.  \\
v^N_k & = &  \left\{
                \begin{array}{cc} 
                                    kB/(n-1),  & \mbox{if \  $k \leq a$},  \\
                                    T \, kB/(n-1), & \mbox{if \  $k > a$},  
                \end{array}
\right.
\end{eqnarray*}
for positive constants $C$ and $B$, $T \geq 1$ and $a \in \{1,2,...,n-1\}$. 
Here we suppose
that a public goods game is repeated a random number of times $\tau \geq 1$, 
with average $\E(\tau) = T$.  Each time each member of the 
group can cooperate at a cost $C$ to itself, resulting in a benefit $B/(n-1)$ to each
one of the other members of its group. 
Defectors incur no costs and produce no benefits.
We suppose that altruists cooperate in the 
first round, and afterwards only cooperate if at least $a$ other members of the 
group cooperated in the previous round. This is a generalization of the well known
tit-for-tat strategy, which corresponds to the case $n=2$, $a=1$.
We will refer to the strategy of the altruists in this example then as 
``many-individuals-tit-for-tat (with threshold $a$)''.

Note that when $T=1$, regardless of the value of $a$, this example is identical to Example 1.
Note also that $\bar v_k = k(B-C)/n$ if $k \leq a$, and 
$\bar v_k = T k(B-C)/n$ if $k > a$.
Again, we suppose that $0 < C < B$, and so both, (C1) and (C2) hold,
since $v^A_1 = -C$ and $v^A_n = (B-C) T$.
It is also easy to see that then (C3), (C5), (C6) and (C7) hold. 

Condition (C4) will only hold under additional assumptions. A very natural
one, that we will assume, unless stated otherwise, is that the threshold $a$ satisfies
\be 
-C + a B/(n-1) \ \geq \  0,
\label{mtft}
\ee
i.e., when altruists keep
playing the game, it is never in their disadvantage to do so.

It is instructive to look into what happens with (C8) in detail: 
$v^A_{k+1} - v^N_k = -C < 0$,
if $k \leq a-1$; $v^A_{k+1} - v^N_k = -CT < 0$ if $k > a$; but in the case $k=a$, 
$v^A_{a+1} - v^N_a = 
T(-C + aB/(n-1)) - aB/(n-1)$. 
Therefore, if (\ref{mtft}) holds as a strict inequality, then
$v^A_{a+1} - v^N_a > 0$, for large $T$, 
and (C8) fails. But if (\ref{mtft}) fails, or holds as an equality, then (C8) holds, for arbitrary $T$.
Note that if (\ref{mtft}) holds as an equality (which can only occur if $(n-1)C/B$ is an integer), then
all the conditions (C1)-(C8) are satisfied.

This model was studied independently in \cite{BR} and in \cite{Joshi}, 
in the trait-group framework.
Both papers identified stable equilibria with positive fractions of altruists (a phenomenon
that can occur only when (C8) fails). But they also observed that altruists could not
invade when rare (a phenomenon that always holds under (C1)). In \cite{BR}
an approach 
was then introduced to provide assortment and allow the gene A to invade when rare.
The authors concluded that such invasion by gene A could only occur under very restrictive
conditions, and that therefore this model and the corresponding notion of 
many-individuals-tit-for-tat were of marginal relevance. One of our contributions 
in the current paper is to rectify this perception. In our framework, we obtain 
values on $m_s$ large enough to indicate that this model should be seriously considered 
as a possible mechanism for the spread of altruistic genes (see Fig. \ref{fig6} and Fig \ref{fig15}). 
Indeed we will see that 
the estimates in \cite{BR} contained an unreasonably pessimistic assumption, that is not 
supported in our framework (see last paragraph in Section 5). 

From a theoretical point of view, this model is still mathematically simple enough to 
lead to an interesting detailed analysis of the conditions under which the gene A can 
spread, in case selection is weak and the group size $n$ is large, with the threshold 
$a$ proportional to $n$. This analysis (illustrated in Fig. \ref{fig16}, Fig. \ref{fig17}, Fig. \ref{fig18}, 
Fig. \ref{fig19} and Fig. \ref{fig20}) will 
be a good illustration of the simplifications that will be obtained, in Section \ref{sec5},
in that regime. 

We see this example as the prototype for an important class of models, that we make explicit in
Example 6 below, and that we believe should be seriously considered and studied.

\

\noindent  {\bf Example 3}. {\it Threshold model:}
\begin{eqnarray*}
v^A_k & = & \left\{
                \begin{array}{lc}
                                    -C,  & \mbox{if \  $k < \theta$},  \\
                                    -C + A, & \mbox{if \  $k \geq \theta$},
                \end{array}
\right.  \\
v^N_k & = &  \left\{
                \begin{array}{lc}
                                     0,  & \mbox{if \  $k < \theta$},  \\
                                     A', & \mbox{if \  $k \geq \theta$},

                \end{array}
\right.
\end{eqnarray*}
for positive constants $C$, $A$ and $A'$, and an integer $\theta \in \{1,2,...,n\}$.
The idea here is simple: the gene A carries a cost, but allows its carriers 
to gain benefits if sufficiently many are in the group. Non-altruists obtain
benefits also when altruists do, but we allowed for the possibility that those 
are smaller or larger than those of the altruists.  

This model may be seen as a simplification of Example 2. It shares with it the 
features that when few altruists are present, they incur costs, but 
when in numbers larger than a threshold, have positive payoffs that may be 
much larger than those costs. The fitnesses in this model are sufficiently 
simpler than those in Example 2, to allow for a more transparent analysis
of its behavior. We will see that this model is of great value when we 
discuss conceptual issues, including Hamilton's rule.
In \cite{Vee1}, the case $n=3$ of this model (called there ``stag hunt game'') 
was discussed in connection to the conceptual issue of the role of Hamilton's rule.
This raised a debate in \cite{Mar} and \cite{Vee2} and further analysis in 
\cite{GWW}. We will comment on this at the end of Section \ref{sec3}.   

Example 3 is also of great 
value for comparison purposes, providing meaningful bounds on the behavior
of more elaborate and realistic models. For instance, the very elaborate 
fitness functions studied in \cite{BGB} are well approximated by those
in Example 3. We are currently reanalyzing the work and ideas from \cite{BGB} in 
the context of our framework, and taking advantage of this relationship
in that project (\cite{SVC2}).

This model can also be seen as a simple instance of another natural class of 
models that we introduce below, in Example 5. 

If $\theta = 1$, then either (C1) or (C2) is violated, since then 
$v^A_1 = v^A_n = -C + A$. So we suppose that $\theta \geq 2$. Under 
this assumption, 
(C1) is immediate from $C>0$, and we suppose that $C < A$,
so that (C2) is also satisfied. 
Conditions (C4) and (C5) are clearly satisfied then. 
Condition (C7) will be satisfied in case $A - C < A'$. 
We have $\bar v_k = -Ck/n$, if $k < \theta$, 
and $\bar v_k = 
A' + (-C + A-A')k/n$, if $k \geq \theta$. So (C3) (and 
therefore also (C6)) may not be satisfied. 
(C3) holds, nevertheless, if $A' \leq A-C$.
(But even under this assumption, (C6) fails, unless $\theta = 2$.)
As for (C8), it fails, regardless of the value of $A'$,
in the same fashion that it failed (in general) in Example 2, since
$v^A_{\theta} - v^N_{\theta-1} = -C + A > 0$.  

Numerical results for Example 3 appear in Fig. \ref{fig7}, Fig. \ref{fig8} and Fig. \ref{fig14}). As with 
Example 2, Example 3 also nice illustrates the simplifications that will be obtained,
in Section \ref{sec5}, when $\delta$ is small and $n$ is large (see (\ref{msapproxex3}) and (\ref{msapproxex3'})).

\

\noindent  {\bf Example 4.} {\it Additive pairwise interactions (general linear fitness functions):}
$$
\begin{array}{llll}
v^A_k & \ = \  a_A + (k-1)a_{AA} + (n-k)a_{AN}  & \ = \   -C + (k-1)B/(n-1)  & \ = \ d_1 + d_2 k, \\
v^N_k & \ = \  a_N + k a_{NA}    +  (n-k-1) a_{NN}  & \ = \          k B'/(n-1)  & \ = \ d_3 k.
\end{array}
$$
Here we suppose that members of the group interact in a pairwise manner throughout their 
lives. Each such pair interaction contributes a certain amount to the 
total payoff of each one  of the two individuals. 
The contribution from each pairwise interaction to each one of the 
two participants depends only on their types. The payoff to a type $i$ interacting with a
type $j$ will be denoted $a_{i,j}$, $i,j$ = A,N. 
In addition, each individual has a self contribution, $a_i$ to its payoff that depends only 
on its type $i$ = A,N. 
These contributions are added to produce the final payoff. The result is displayed above,
and then rewritten in terms of $C= -a_A - (n-1) a_{AN}$, $B/(n-1) = a_{AA} - a_{AN}$, 
$B'/(n-1) = a_{NA} - a_{NN}$, where we incorporated the assumption that 
$$
a_N + (n-1) a_{NN} = 0.
$$
This assumption carries no loss of generality, since a constant can be added to all the 
payoffs $v^A_k$ and $v^N_k$ without modifying the behavior of our process. 
(The behavior of the process is clearly not modified by multiplying all fitnesses, 
$w^A_k = 1 + \delta v^A_k$, $w^N_k = 1 + \delta v^N_k$, by the
same constant. If we add a constant $v$ to all the payoff functions $v^A_k$, $v^N_k$, 
the new fitnesses are equivalent in this sense to the old fitnesses
with $\delta$ replaced by $\delta/(1+\delta v)$.)
This condition amounts simply to our convention that $v^N_0 = 0$. 
Expressing the fitness functions 
in terms of $C$, $B$ and $B'$, 
makes their relationship with Example 1 and Example 5
below easy to
see.  Finally the result is also rewritten in an equivalent form that emphasizes 
the nature of the dependence of the fitnesses on $k$: they are linear functions. 
Here $d_1 = -C - B/(n-1)$, $d_2 = B/(n-1)$ and $d_3 = B'/(n-1)$. 

An important point to make is that given linear fitness functions 
$v^A_k$ and $v^N_k$, with $v^N_0 = 0$, as above, they can always be represented in
the other two ways, with an appropriate choice of the constants.
For instance, we can take $C = -d_1 - d_2$, $B/(n-1) = d_2$ and $B'/(n-1) = d_3$, 
and then take
$a_A = a_N = a_{NN} = 0$, $a_{AA} = (B-C)/(n-1)$, $a_{AN} = -C/(n-1)$, $a_{NA} = B'/(n-1)$. 

When $n=2$, the current example is the most general possible choice of the payoff functions
$v^A_k$ and $v^N_k$. But this is obviously not the case when $n \geq 3$. For each value of 
$n$, the most general form of the payoff functions are polynomials of degree $n-1$. 

The linearity of the functions $v^A_k$ and $v^N_k$ will play important roles in 
relating our results to other concepts, especially Hamilton's rule. This is one of the 
reasons this is a major class of models. The mathematical equivalence between these
linearities and having pairwise additive interactions should not confuse one into 
thinking that the linearities imply that the fitnesses must indeed have
originated from that special
kind of interaction. 
In Example 1, the members
of a group may be interacting in a collective way (hunting together, warfare, etc). 
All that the mathematical equivalence says is that the fitnesses obtained there, are the
same ones of a, ficticious in this case, pairwise interaction scenario. 
This is a good illustration of two formally equivalent, but materially different models.

There are realistic stories that are conceptually associated to the public goods game, 
Example 1, but lead to the more general payoff in the current example, with $B' \not = B$.
For instance, the altruistic activity could be hunting more dangerous but also more 
nutritional prey. If the products of the hunts are always shared by the group, we 
have the model in Example 1. But if the hunters are able to consume the best 
part of the hunt, before sharing the rest with the group, we would have $B' < B$. 

While materially different from the public goods game, Example 1, the current example is 
formally equivalent to it when the following equivalent conditions hold:
\be
B= B', \ \ \ \ \ 
a_{AA} - a_{AN} \ = \ a_{NA} - a_{NN}, 
\ \ \ \ \
a_{AA} - a_{NA} \ = \ a_{AN} - a_{NN}.
\label{egfs}
\ee
Condition (\ref{egfs}) is know as ``equal gains from switching'', since the payoffs in the $2 \times 2$
matrix $a_{i,j}$, $i,j = $ A, N, change by the same amounts 
if one switches strategies, regardless of what the other player is doing. 
Unfortunately the terminology ``additivity condition'' or ``linearity condition''
is also used in the literature 
for (\ref{egfs}). This is confusing, since in our context, additivity refers to the 
fact that the payoffs $v^A_k$ and $v^N_k$ are obtained additively over the $k-1$ pairwise 
interactions that each individual has with the other members of its group. This additivity
has no relationship with (\ref{egfs}). And in our context, linearity, refers to the 
linearity of $v^A_k$ and $v^N_k$ as functions of $k$. As we explained above, in the 
mathematically standard way in which we are using the terms pairwise additivity and linearity, they are 
equivalent to each other, and logically independent of (\ref{egfs}). 

Under condition (\ref{egfs}), it is common to use the representation 
$a_{NN} = 0$, $a_{AA} = -c + b$, $a_{AN} = - c$, $a_{NA} = b$.
When $b > c$, this is a classical prisoner's dilemma. It corresponds to 
$C = (n-1)c$, $B = B' = (n-1)b$. 

The matrix $a_{i,j}$ represents the lifetime payoff for each pairwise interaction. 
This lifetime payoff may result from the accumulation of payoffs from iterated games.
In this way we can see that the setup in this example 
is flexible enough to accommodate a gene A that produces a conditional behavior 
over such iterated games, like, for instance a tit-for-tat strategy. For this, suppose 
that each pair of individuals interact repeatedly with payoffs given by the 
standard prisoner's dilemma matrix. If type N always defects, and type A uses a 
tit-for-tat strategy, we have $a_{NN} = 0$, $a_{AA} =(-c + b)\, T$, 
$a_{AN} = - c$, $a_{NA} = b$, where $T$ is the average number of repetitions of
the basic interaction over a lifetime. If $T > 1$, (\ref{egfs}) fails, and we have 

\be
v^A_k = -(n-1)c + ((b-c)T + c) (k-1), \ \ \ \ \ v^N_k = bk.
\label{ipd}
\ee

It is common to write $B-B' = D$ and call it a ``synergy'' term. It represents an additional
benefit (possibly negative) to altruists when interacting with other altruists. In the 
iterated pairwise prisoner dilemma game with A playing tit-for-tat, (\ref{ipd}), we have
$D = (b-c)(T-1)(n-1)$. 

Deciding when each one of the conditions (C1)-(C8) holds in 
the current example is tedious and not so relevant. We observe only a few facts. 
Assuming $0<C<B$s (C1), (C2) and (C4), that only depend on $v^A_k$. 
If $B'> 0$, then also (C5) holds. The other conditions depend on how $B'$ relates
to $B$ and $C$. We just make the simple remark that if $0 < C < B$ and $B'$ 
is close enough to $B$, then all the conditions (C1)-(C8) hold, since they hold 
with slack when $0<C<B=B'$ (Example 1). 

In Example 1, we observed that the assumed  
linearity of $v^A_k$ and $v^N_k$ there may not be realistic. The same observation holds 
about pairwise additivity of interactions. Many interactions in a group are between
pairs, but it is not always clear that their effects on fitness should be additive. When 
an individual interacts repeatedly with another member of the group, like in the 
story that lead to (\ref{ipd}), it may not be able to interact as often with the 
other members of the group. Also, the beneficial effects of the pairwise interactions 
may saturate, and be sub-additive, rather than additive. Additivity/linearity is
mathematically a natural first level simplification/approximation. But one should 
be aware of its limitations. With this in mind, we turn to the next example.

\

\noindent  {\bf Example 5.} {\it Variable costs and benefits:}
\begin{eqnarray*}
v^A_k & = & -C_k + (k-1)B_k/(n-1),  \\
v^N_k & = &    k B'_k/(n-1),
\end{eqnarray*}
Remarks in Example 1 and 4, above, motivate this class of of examples.
Without further assumptions on the costs and benefits functions, 
$C_k$, $B_k$ and $B'_k$, any model can 
be fit into this form. So that what we are proposing here is first a
convenient notation for comparing models. Next we discuss some 
interesting assumptions on the costs and benefits functions.

It is very natural, in various applications, to assume that $C_k$ is non-increasing 
and $B_k$ and $B'_k$ are non-decreasing.
If the gene A prompts its carriers to act in some collective way, it is often the 
case that the cost to each participant decreases with the number of participants, while 
the total benefits produced grow faster than linearly with the number of participants.
This is called an increasing return to scale. Reasonable assumptions can be 
that $C_k$ decrease as a power of $k$, $C_k = C/k^{a_1}$, for some constant $a_1>0$, 
while $B_k = a_2 k^{a_3}$ and $B'_k=a_4 k^{a_5}$, with $a_2 > 0$, $0 < a_3 < 1$, 
$a_4 > 0$, $0 < a_5 < 1$. 

Another distinct assumption on the benefit functions is that 
$(k-1) B_k / (n-1)$ and $k B'_k/(n-1)$ 
first grow slowly with $k$, then steeply (close to a threshold value of $k$)
and then more slowly again, as the gains from scale saturate. 
For instance
this will happen if 
$B_k = b k / (1 + d k^2)$, $B'_k = b' k / (1 + d' k^2)$,
with positive constants $b$, $d$, $b'$, $d'$.

An interesting class of models covered by the current example is 
the object of \cite{HMND}. Experimental results with microbes often indicate 
the need to consider non-linear payoff functions, as those discussed in 
the current example; see, for instance \cite{CRL} and \cite{SDZ}. 

In case of a collective 
action that requires a minimum number $\theta$ of participants, we should have 
$B_k = B'_k = 0$, for $k < \theta$. And for larger values of $k$, $B_k$ and 
$B'_k$ should grow, but again typically not linearly. For instance, 
gene A could promote a behavior that
can only be implemented in groups of at least 4 individuals, say. This could be a type of
large game hunt, that requires 4 hunters. Gene A causes
changes to the individual's phenotype that make this kind of hunt possible, but at the expense
of adding expensive muscular and/or brain tissue.
Types N just hunt individually small game. We suppose that the hunters share their 
product with the group, and that large game produces greater benefits per person
in the group, than small game hunt. How will $B_k$ and $B'_k$ grow when $k \geq 4$?
The answer will depend on ecological conditions, and detailed aspects of the hunting 
technique.
Can several different groups
hunt simultaneously? Would the hunt be more efficient with 7 hunters than with 
4? If most group members are hunting large game, would the productivity of small 
game hunt increase so as to make it advantageous for the group to combine both 
types of hunt? And so on. 

\

\noindent {\bf Example 6.}  {\it Iterated game. Altruists cooperate conditionally, based on feedback:}
\begin{eqnarray*}
v^A_k & = & T_k \, (-C_k + (k-1)B_k/(n-1)),  \\
v^N_k & = & T_k \,  k B'_k/(n-1).
\end{eqnarray*}
Here $T_k$ can be seen as an average number of repetitions of a basic activity.
This class of models builds on the models in Example 5 in a fashion that generalizes
the way Example 2 was built on Example 1. We are supposing that a certain 
activity presets itself to the group periodically. The output each time depends on 
the behavior of the group members, and gene A modifies this behavior. Carriers of 
gene A behave first in a way that is beneficial to the group. And afterwards, they 
will or not continue acting in this way, depending on feedback that they receive. 
For instance, if the activity is a type of collective hunt, they will have feedback 
as they consume the product of the hunt. In Example 2, the feedback was a count of
the number of participants, and this is also a possibility, but not the only one.
In Example 2, $T_k$ was a step function, jumping from $1$ to $T$, at $k = a+1$.
But it seems also natural to consider smoother functions $T_k$, that increase first 
slowly, then steeply, and then saturate. This could result from the fact that the 
feedback, from each repetition of the activity, is subject to random noise, and 
only gives clear cues to the altruists, outside of a critical window of values of $k$.
For values of $k$ in the critical window, altruists may repeat the activity a few 
times, before deciding to stop participating.



Models represented in Example 5, with non-increasing $C_k$ and non-decreasing $B_k$,
have a natural threshold value of $k$, where their payoff to altruists becomes 
positive. In Example 1, this threshold value $k=a+1$, corresponds to the condition (\ref{mtft})
that appears then in Example 2. If the feedback that affects the willingness of 
altruist to continue participating in the activities promotes a function $T_k$ 
that (as in Example 2, with assumption (\ref{mtft}))
starts growing above that threshold, the resulting behavior becomes more 
adaptive than it would be without feedback effect. 

Mathematically the models in the current example can be incorporated in Example 5,
by modifying the definition of $C_k$, $B_k$ and $B'_k$. But the point of the 
current example is to provide justification for, and a mechanism behind,
a class of models in which 
as $k$ increases, $v^A_k$ switches from small negative values to much larger 
positive values after a threshold value of $k$ is crossed. 
These models can be approximated by  and/or compared to the simple case 
given in Example 3. 
It then becomes natural to analyze that 
model with ratios $A/C$ which can be as large as 100, or 1000. For instance, 
suppose the species under consideration to be early humans, with an adult 
reproductive lifespan of over 20 years. If the activity in consideration is 
repeated with a frequency of 50 per year, and if with negative feedback altruists
stop participating after about 10 repetitions, then we can consider a factor
of $50 \times 20 / 10 = 100$ between $T_k$ with large $k$ and $T_k$ with small $k$. 

Obviously the perceived feedback cannot be the
payoff in the model, which is related to expected number of offspring in
the future. But it is sufficient that the feedback be strongly correlated with this payoff.
This is not a special problem about altruism, when behavior is mediated 
by feedback, natural selection will align reaction to feedback with fitness. 

The combination of increasing returns to scale, with the possibility of only pursuing the
behavior when it is advantageous to the actors, due to a high level of collective
participation, may have been a powerful set of mechanisms that led to the evolution
of altruistic and cooperative social traits. This idea has, for instance, 
been explored in \cite{BGB}, 
where a model of that nature was analyzed in connection with the trait
of punishing, at a cost, those that do not cooperate with the group.
In that model, types A produce first a costly signal announcing that they are willing
to participate in costly punishment. But they only implement the punishment if 
a sufficiently large number of group members signaled their willingness of 
doing it too. 

Another story that fits with increased returns to scale, modified by 
feedback induced discontinuation, can be suggested for the emergence of
``compassionate feelings'' for group members.
A gene that promotes those emotions, would cause
their carriers to help fellow group members in need, at a cost to themselves.
While there are few group members carrying this gene, they will not be able
to do much for all those that may need help, in a large group.
Feedback, in the form of frustration for not being able to help as they want,
may lead them to discontinue their helping activities, in a short while. 
But with enough carriers of that gene in a group, they become able to 
successfully provide help to all that need it. Under this condition they 
pursue this activity, throught their lives. 
At the same time, under this conditions, the altruists themselves have their 
fitnesses increased substantially, from living in a helping environment.
This suggests that
this story may be represented by a payoff function of the type that we are introducing
in the current example, with $T_k$ that can reasonably be taken to vary by a factor of 100
or more, as a threshold window $k$ is crossed.


\section{Conceptual discussion}
\label{sec3}

We start next on a long conceptual discussion of our results, and how they
may help clarify some controversial issues related to the emergence of
altruism through natural selection.

First we consider how our results and the underlying mechanisms revealed
by them relate to ``group selection'', ``multi-level selection''
and ``kin selection''. This is not the
place, and neither do we have the expertise to discuss in detail the
various nuances of the semantics involved in these questions. But a few
words are in order, and should be of value to the readers.

The use of the expression ``group selection'' has changed over the years, and 
is still somewhat controversial. Nevertheless it seems to us that under any reasonable 
use of that expression, group selection is an important force operating in our 
setting. In our framework, individuals belong to groups, 
groups compete among themselves, and in this way the fitness
of individuals depends on the constitution of their groups. In our 
typical examples, the fitness of each altruist is strictly smaller
than the fitness of the non-altruists in its group 
(Condition (C7)).
It is only through
the higher fitness of groups with many altruists, that the altruistic
gene A is then able to survive and spread.
Individuals also compete for reproduction with other members of their 
own group, so that in our setting group selection is one of the 
components of a ``multi-level selection'' process. 

Is the mechanism of kin selection present when the mutant A spreads?
We understand kin 
selection as a process in which copies of a gene,
originating from a recent common ancestor,
interact with each other providing themselves with an average fitness large
enough for this gene to survive and spread. This is precisely what happens
in our case, and is also what we tried to capture in more detail in the
viability criterion and associated survival mechanism presented at the end of Section 1. 
In other cases, the assortment and organization
of the genes could be caused by kin recognition. In our case it is caused
by viscosity, that results from the group structure of the population
and limited migration. In a model in which an isolated mutant has
relative fitness lower than the wild type (condition (C1) in our case),
its only hope for survival is in the creation by chance of a few of its copies,
that happen to be so arranged that they have 
higher average fitness than the wild type in the population at large,
and for the structure of the population to be such that this gene can 
then spread by what we called survival mechanism. 

In this connection,
condition (\ref{rho>1nu}) can be seen
as providing a ``gene's eye view'' of viability in our setting.
As noted above, the left hand side in this condition is simply 
the average fitness of the gene, when it is arranged in the best 
possible stable way, to assure its spreading. It is common to 
refer to the average fitness of a gene A as its neighbor modulated
fitness. Under certain conditions it is known that the neighbor 
modulated fitness can be computed also as an ``inclusive fitness'',
in which one adds the effects of a randomly selected gene A from 
the population on the other genes A. At this point, we are not 
sure how far this method can be extended. 
(See \cite{Gra2} for a general investigation of this question.)
We will address later in 
this paper the issue of the validity of the related Hamilton 
rule in our setting, but we will defer the answer to the question of when the 
neighbor modulated fitness of the gene A can also be computed as 
an inclusive fitness to a later investigation.  

It is worth also clarifying that in our view multi-level selection and kin
selection are not the same concept, even if both are central to
our study. We see multi-level selection as a process in which the demographics
and/or the biology (including behavior)
associates individuals to groups in such a way that the reproductive success of 
each individual depends on the composition of its group. 
Kin selection can happen, as
in our framework, in a multi-level selection setting. But it can also
happen in populations that are organized in other ways, in which 
individuals are not sorted into groups. Moreover, in our framework, in the
late stage, when types A exist in numbers comparable with $g$,
multi-level selection will continue to be  a basic driving force acting on the
population. But the average fitness of types A will no longer result only
from their interaction with other types A that are close kin. Whether
one should still refer to kin selection as an important force then is an issue
that we postpone to the paper in preparation in which we study the late
stage (\cite{SVC1}).

In the way that we use the expressions ``multi-level selection'' and ``kin selection''
in the discussion above, they are qualitative concepts, rather than computational
or accounting procedures.
These concepts are nevertheless sometimes associated to certain
computational procedures:
Multi-level selection is sometimes associated to the Price
equation, and kin selection is often associated to either neighbor
modulated fitness, or inclusive fitness computations. But specially in an
area in which semantic issues are a source of difficulties, one should
carefully separate concepts from computational procedures.
As we explained above, while the concept of neighbor modulated fitness
fits easily into our framework, it is currently not clear to us that
the concept of inclusive fitness could fit as well. The Price equation
obviously applies in our framework, since it requires minimal conditions,
and is a natural, mathematically rigorous,
tool to consider when studying group-structured populations.
We will elaborate below on what it adds to the solution 
of the specific problems that we are addressing in this paper, 
and why we did not use it in our analysis in Section 1.

To decide on the viability of the mutant A in our framework, we see no
simpler mathematical method (when selection is strong -- we will study
the important simplifications in case of weak selection later in this
paper) than computing $\rho$ or $\nu$. This does not mean that alternative
or complementary methods cannot add insights, intuition and relevant 
information. Moreover, it is important to understand how different 
approaches relate to each other, and computational parameters relate to
experimentally accessible variables.
With this in mind, 
we turn now to a discussion of how 
our mathematical approach compares to the use of the Price equation, and 
to the role of Hamilton's rule in our setting.

Before proceeding we need to introduce some more notation:
$$
N_0(t) = g - (N_1(t) + ... + N_n(t)),  \ \ \ \ \ 
f(t) = (N_0(t)/g, N_1(t)/g, ... , N_n(t)/g),
$$
$$
p(t) = \frac{N^A(t)}{gn} = \sum_k f_k(t) (k/n).  
$$
This means that $f(t)$ is the distribution of the various types of groups, including 
groups of type 0, in generation $t$, 
and $p(t)$ is the fraction of altruists in the population then. 


The Price equation provides the expected value of $p(t+1)$, when $f(t)$ is given. 
It can be stated in several mathematically equivalent versions. 
In our setting the simplest one is
\be
\E(p(t+1)\,|\,f(t)) \ = \ p(t) \, \frac{W^A(f(t))}{W(f(t))},
\label{price}
\ee
where 
$$
W(f) = \sum_k \bar w_k f_k, 
\ \ \ \ \ \ \ \ 
W^A(f) = \frac{\sum_k w^A_k k f_k}{\sum_k k f_k},
$$
are, respectively, the average fitness of the individuals in the population 
and the average fitness of the altruists in the population, when the distribution
of the group types is given by $f=(f_0,f_1,...,f_n)$. 
Equation (\ref{price}) is an immediate consequence of the fact that each individual 
has an expected number of offspring proportional to its relative fitness. By adding and 
subtracting terms, it can be rewritten as
\be
\begin{array}{lcl}
W(f(t)) \, \E (p(t+1)&-&p(t) | f(t)) \ = \ p(t) ( W^A(f(t)) - W(f(t))) \\
& = & \ \sum_{k} (w^A_k - \bar w_k) (k/n) f_k(t) \ 
+ \ \sum_{k} (\bar w_k - W(f(t))) (k/n) f_k(t) \\
& = & \ \E ((w^A_K - \bar w_K) (K/n)) \ + \ \mbox{Cov} (\bar w_K , K/n),
\end{array}
\label{price++}
\ee
where $K$ is a random variable with $\P(K=k) = f_k(t)$. 
In the E.S., since the fraction of altruists is 
negligible in the large population, we have in good approximation $W(f(t)) = 1$,
reducing (\ref{price}) to
\be 
\E(p(t+1)\,|\,f(t)) \ = \ p(t) \, W^A(f(t)).
\label{price+}
\ee
Can (\ref{price}) or the equivalent (\ref{price++}), or the simplified (\ref{price+})
be used to predict when altruism can survive in our setting? 
The answer is negative. 
In generation $t=0$ we have $f(0)=(1-1/g,1/g,0,0,...,0)$, and the average fitness of 
A, which is simply the fitness of the only A present, is $W^A(f(0)) = w^A_1 < 1$,
under condition (C1). Therefore (\ref{price+}) leads to 
$\E p(1) = w^A_1 / g < 1/g = p(0) $ 
for any value of $m$, as we already new.  
If one wants to iterate (\ref{price+}) over time, in order to learn under 
what conditions $\E p(t)$ eventually increases and does not vanish, one needs
to be able to compute $f(1)$, $f(2)$, ... etc. The well known problem is that 
(\ref{price+}) does not provide information about $f(t+1)$ given $f(t)$. 
Actually (\ref{price+}) carries no information at all about $m$ and how the 
groups are formed in generation $t+1$. 

The Price equation in its various forms is a useful 
tool for many purposes. When the right hand side is split, as in (\ref{price++}), 
into two terms that correspond respectively to intergroup and intragroup 
competition, it carries great heuristic power, and beauty. We hope that, nevertheless, 
our current study may help clarify some of its limitations in the analysis of 
evolution in group structured populations.
It is interesting to look into what (\ref{price+}) tells us in combination 
with what we already know about the evolution of our population. When the gene
A survives, in the S.E.S., (\ref{ses})
implies that $(f_1(t),f_2(t),...,f_n(t)) = C(t) \nu$, 
where $C(t)$ is time dependent and random, but one-dimensional. Therefore 
$W^A(f(t)) = (\sum_k w^A_k k \nu_k) / (\sum_k k \nu_k) = \rho$, thanks to
(\ref{rhonu'}). We obtain now from (\ref{price+})
$$
\E(p(t+1)\,|\,f(t)) \ = \ \rho \, p(t). 
$$
This is compatible with the growth rate given by (\ref{ENrhot}) and
(\ref{ses}), and is not 
new information to us, but the consistency is reassuring. 
When $\rho > 1$, $\E(p(t))$ first decreases, but then eventually increases, reaching
growth rate $\rho$ in the S.E.S.. 
As explained above, 
in this case either the gene A dies out early on, or else, natural selection 
organizes it in groups according to the distribution $\nu$, that is stationary 
for the evolution driven by $M(A+B)$ and maximizes the rate of growth of $p(t)$
among all such stationary distributions.
The power of (\ref{NM}) over (\ref{price+}), is that it provides the
evolution of the whole distribution over group types, not just the fraction of
altruists in the population.


\ 

Next we address the question of whether Hamilton's rule applies in our setting,
to provide the condition under which altruism, 
and other genetically determined behaviors that are costly to the actor,
can spread. The answer here greatly
depends on what one means by ``Hamilton's rule''. In its broadest sense, it is 
natural to use this title for any inequality that is a necessary and sufficient 
condition for altruism to have a positive probability of spreading, in other 
words, for any ``viability condition''. With this 
interpretation, (\ref{rho>1nu}) 
is a generalized Hamilton rule that is universal in our framework, when started with a 
single altruist. In this paper we will use the terminology ``generalized Hamilton rule'', 
or ``generalized Hamilton condition'',  
and ``viability condition'' with equivalent meanings (some may prefer ``survivability 
condition''). We hope that rather than 
creating confusion, this usage will shift the discussion from a semantic matter
into questions with scientific content: How does the viability condition 
(\ref{rho>1nu}) simplify in special cases? Can it be formulated in terms 
of concepts of relatedness, costs and benefits? Can it be formulated in 
terms of experimentally meaningful variables? Can it be written in ways 
that help compare different models?

We should clarify that here we are considering the question of when 
the altruistic gene A is viable, starting from a single copy of it. 
Conditions for selection to favor gene A when the process is started 
with a number of copies of A comparable to $g$ are a different problem, 
that we will address when studying the evolution in our framework in 
the late stages (\cite{SVC1}). Here we will only address this further question in the
special important case of pairwise additive interactions, Example 4, so 
as to illustrate how things can change when the gene A becomes common in
the population.


Before we can analyze how (\ref{rho>1nu}) compares in spirit and content with 
more traditional forms of Hamilton's rule,
we need to introduce and review several additional concepts.


Suppose that in generation $t$,
a group is chosen at random, and its members are ordered in a random fashion. 
The chosen group is called the focal group, 
the first individual in the ordering is called the focal individual and the second individual 
in the ordering is called the co-focal individual. Note that this random experiment 
is equivalent to choosing a 
focal individual at random and then choosing a co-focal individual at random from the 
other $n-1$ individuals in the focal's group. 
We will denote by $\P_t$ probabilities that refer to the sampling
just described, and use $\E_t$ for corresponding expected values.
We denote by $A_j$ the event that the j$th$ 
individual in the random ordering of the members of the focal group is a type A. And 
we denote by $I_j$ the indicator of the event $A_j$, i.e., $I_j$ is the random variable
that takes value 1 if $A_j$ occurs and value 0 if its complement, $A_j^c$, occurs. 
We denote by $K = I_1 + ... + I_n$ the random number of types A in the focal group. 
And we set $\hat p = K/n$, for the random fraction of altruists in the focal group.

Clearly $\P_t(A_j) = p(t)$ does not depend on $j$ and is the fraction of altruists in 
generation $t$. Linearity of expectations yields the following relationships, that 
will be of great use:

\be
\E_t(K) \ = \ \E_t(I_1)+...+\E_t(I_{n})  \ = \
n \P_t(A_1) \ = \ n p(t), 
\ \ \ \ \  \E_t(\hat p) = p(t).
\label{KPA}
\ee
\be
\E_t(K-1|A_1) \ = \ \E_t(I_2|A_1)+...+\E_t(I_n|A_1)  \ = \
(n-1) \P_t(A_2|A_1).
\label{KPAA}
\ee
\be
\E_t(K|A^c_1) \ = \ \E_t(I_2|A^c_1)+...+\E_t(I_n|A^c_1)  \ = \
(n-1) \P_t(A_2|A^c_1).
\label{KPAAc}
\ee


For $t$ in the S.E.S., we have the following fundamental relationship
\be
\P_t(K=k|A_1) \ = \ 
\frac{k \nu_k}{\sum_{k'}k'\nu_{k'}} 
\ = \ \Ps(K=k),
\label{PASES}
\ee
for $k \geq 1$, 
where the second equality introduces a new notation.
To see this, first note that conditioning on $A_1$ implies that the gene A has survived into 
the S.E.S., and therefore (\ref{ses}) holds.
The sampling, at time $t$, is therefore of a population that has mostly groups with
no altruists, but has also a large number (of order $\rho^t$) of groups with altruists,
distributed according to $\nu$. If the conditioning was on $K \geq 1$, rather than on
$A_1$, the conditional probability would be $\P_t(K=k|K\geq1) \, = \, \nu_k$. We would
be just sampling unbiasedly from the groups with altruists. But conditioning on $A_1$ 
introduces size bias: For $k \geq 1$, 

\begin{eqnarray*}
\P_t(K=k|A_1) 
\ & = & \
\frac{\P_t(A_1|K=k) \, \P_t(K=k)}{ \sum_{k'} \, \P_t(A_1|K=k') \, \P_t(K=k')}    
\\
\ & = & \
\frac{(k/n) \, \nu_k \, \P_t(K\geq1)}{ \sum_{k'} \, (k'/n) \, \nu_{k'} \, \P_t(K\geq1)}
\ = \ 
\frac{k \nu_k}{\sum_{k'}k'\nu_{k'}}.
\label{sizebias}
\end{eqnarray*}

Using (\ref{PASES}), we can now rewrite our viability, or generalized Hamilton rule, (\ref{rho>1nu}) as
$$
\frac{\sum_k w^A_k k \nu_k}{\sum_k k \nu_k}
\ = \  \sum_k \, w^A_k \, \Ps(K=k) \ = \ \Es(w^A_K) = \rho \ > \ 1,
$$
which for arbitrary strength of selection $\delta > 0$, is equivalent to
\be
\frac{\sum_k v^A_k k \nu_k}{\sum_k k \nu_k} \ = \
\sum_k \, v^A_k \, \Ps(K=k) \ = \ \Es(v^A_K ) \ > \ 0.
\label{HR+}
\ee
Conditioning on $A_1$ means that the focal individual is a randomly
chosen altruist from the population. Inequality (\ref{HR+}) states
that the expected payoff to this individual, from its behavior and
the behavior of the others in its group, is positive.
(This expected value can be seen as a ``neighbor modulated'' expectation.)
This starts to look more like a traditional Hamilton rule. But one should keep
in mind that this is only valid because we assumed the sampling to
be done during the S.E.S.. Had we sampled from the
initial generation, we would have obtained $\E_0(v^A_K | A_1) = v^A_1 < 0$,
by (C1), regardless of the value of $m$.
Still, with this caveat, (\ref{HR+}) carries heuristic
power, and adds more meaning to (\ref{rho>1nu}).
For computational purposes, though, one cannot use
(\ref{HR+}) until one has information on $\nu$.

In order to further pursue
the relationship between (\ref{HR+}) and the
expressions traditionally known as Hamilton's rule,
we need to consider special examples.
And before we can do it, we need to introduce the relatedness into our framework.

If we define relatedness $r_t$ as the regression coefficient of $I_2$ on $I_1$ (or equivalently, 
of $A_2$ on $A_1$, or of $A^c_2$ on $A^c_1$), we have 
\begin{eqnarray*}
r_t \  & = & \
\frac{\mbox{Cov}_t(I_1,I_2)}{\mbox{Var}_t(I_1)} \ = \
\frac{\P_t(A_1 A_2) - (p(t))^2}{p(t)(1-p(t))} 
\  = \
\frac{\P_t(A_2|A_1) - p(t)}{1-p(t)}  \\
\ & = & \ \frac{\mbox{Cov}_t(1-I_1,1-I_2)}{\mbox{Var}_t(1-I_1)}
\ = \
\frac{\P_t(A^c_1 A^c_2) - (1-p(t))^2}{p(t)(1-p(t))}
\  = \
\frac{\P_t(A^c_2|A^c_1) - (1-p(t))}{p(t)}.
\end{eqnarray*}
Equivalently, $r_t$ can be defined by each one of the following four identities:
\be 
\begin{array}{lcl}
\P_t(A_2|A_1) \  = \ r_t + (1-r_t)p(t),  & \ \ \ \ \  & \P_t(A^c_2|A_1) \  =  \ (1-r_t)(1-p(t)), \\
\P_t(A_2|A^c_1) \  = \ (1-r_t)p(t), & \ \ \ \ \  & \P_t(A^c_2|A^c_1)  \ = \ r_t + (1-r_t)(1-p(t)).
\end{array}
\label{PAA}
\ee
In particular, 
\be 
r_t \ = \ \P_t(A_2|A_1) - \P_t(A_2|A_1^c).
\label{rP-P}
\ee
Combining (\ref{rP-P}) with (\ref{KPAA}) and (\ref{KPAAc}) yields 
\be
r_t \ = \  \frac{\E(K-1|A_1) \, - \, \E(K|A_1^c)}{n-1} \ = \ 
\frac{\E(K|A_1)  \, - \, \E(K|A_1^c) \, - \, 1}{n-1}.
\label{rtK}
\ee

The relatedness $r_t$ is closely associated to 
Wright's $F_{ST}$ statistics, defined as: 
$$
F_{ST,t} \ = \ \frac{\mbox{Var}_t(\hat p)}{\mbox{Var}_t(\hat p) \ + \ \E_t(\hat p (1 - \hat p))}.
$$
(The numerator measures intergroup variability, and the second term in the denominator measures 
average intragroup variability.)
To see how $F_{ST,t}$ relates to $r_t$, we write
\begin{eqnarray*}
\mbox{Var}_t(K) \ & = & \ \mbox{Var}_t(I_1 + ... + I_n) \ = \ 
n \, \mbox{Var}_t(I_1) \ + \ n(n-1) \, \mbox{Cov}_t(I_1,I_2)                 \\
\ & = & \
n p(t)(1-p(t)) \ + \ n(n-1) p(t)(1-p(t)) r_t                                 \\
\ &  = & \ n p(t)(1-p(t)) \, (1 \, + \, (n-1)r_t),
\end{eqnarray*}
and
\begin{eqnarray*}
\mbox{Var}_t(\hat p) \ + \ \E_t(\hat p (1 - \hat p)) \ & = & \ 
\E_t( (\hat p)^2) \, - \, (\E_t(\hat p))^2 \, + \, \E_t(\hat p) \, - \, \E_t( (\hat p)^2)  \\
\ & = & \ 
-(p(t))^2 \, + \, p(t) \ = \ p(t)(1-p(t)).
\end{eqnarray*}
Since $\hat p = K/n$, we obtain,
\be
F_{ST,t} \ = \ \frac{1 \, + \, (n-1) \, r_t}{n},
\ \ \ \ \ \mbox{or, equivalently,} \ \ \ \ \
r_t \ = \ \frac{n \, F_{ST,t} - 1}{n-1}.
\label{FSTr}
\ee
In particular, when $n$ is large $F_{ST,t}$ is close to $r_t$. 
A by-product of the computations above is the identity
$$
F_{ST,t} \ = \ \E_t(\hat p | A_1) \ - \ \E_t(\hat p | A^c_1),
$$
which follows from 
comparing (\ref{FSTr}) with (\ref{rtK}). 
This identity can also be rewritten as
$$
F_{ST,t} \ = \ \frac{\E_t(\hat p I_1)}{p(t)} \ - \ \frac{\E_t(\hat p (1- I_1))}{1-p(t)}
\ = \ \frac{\mbox{Cov}_t (\hat p, I_1)}{p(t)(1-p(t))}
\ = \ \frac{\E_t(\hat p I_1) - p(t)}{1-p(t)},
$$
showing the $F_{ST,t}$ is the regression coefficient of $\hat p$ on $I_1$. 

\begin{figure}[th!]
\begin{center}
\includegraphics[width=6cm]{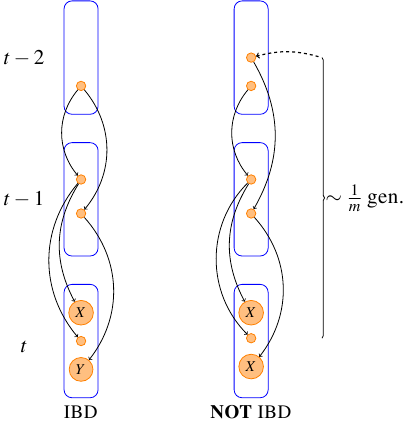}
\end{center}
\caption{This diagram illustrates the concept of identity by descent (IBD) as it is employed in the framework we have 
introduced.
 Two individuals  $X$ an $Y$  in a given group in generation $t$, regardless of their type, are identical by descent (IBD) if their 
lineages, when followed back in time, coalesce before a migration event (indicated by 
a dashed arrow in the figure in the right panel). Considering a migration rate of $m$, migration 
typically takes place within a random number, of order $1/m$ of generations back. }
\label{fig9}
\end{figure}

Our framework allows for a very natural definition of ``kin'' and ``genetical identity by 
descent'', IBD for short (see Fig. \ref{fig9}). 
Say that two individuals in the same generation are $l$-kin in case they share 
a common ancestor at most $l$ generations in their past. (The 1-kin of an individual 
are its siblings, its 2-kin are its siblings and cousins, etc.) Because the number of 
groups $g$ is large, individuals from different groups have negligible probability of 
being $l$-kin, unless $l$ is comparable to $g$. As for individuals in the same group,
migration events in their lineages play a major role in their being or not kin. 
If, when we follow their lineages back in time, we find a migration event in one of
these lineages before they coalesce, then we know that the probability that they 
will coalesce in a time that is much shorter than $g$ is negligible. With this
in mind, say that two individuals in the same generation are IBD in case following 
their lineages backwards in time, they coalesce before (in the backwards sense)
a migration event happens in either one. From the discussion above, we know that 
being IBD is essentially the same as being $l$-kin for some $l$ that may have to
be large, but is not comparable to $g$. 

We denote by $D$ the event that the focal and the co-focal individuals are IBD, and
define genetic relatedness for the allele A by
$$
R_t \ = \ \P_t(D|A_1).
$$ 
In the E.S., $\P_t(A_1) = \P_t(A_2) = p(t)$ is negligible, and so is also 
$\P_t(A_1 A_2 D^c)$ (we adopt the common convention of omitting the intersection symbol $\cap$).
This implies that
$
r_t  \, = \, \P_t(A_2|A_1) \, = \, \P_t(A_2 D |A_1) + \P_t(A_2 D^c |A_1) \, = \, \P_t( D |A_1) \,  = \, R_t.
$
Using also (\ref{KPAA}), we have then
$$
R_t \ = \ r_t \ = \ 
\P_t(A_2|A_1)
\ = \ \frac{\E_t(K-1|A_1)}{n-1}, 
\ \ \ \ \ \ \ \ \ 
\mbox{for $t$ in the E.S.}. 
$$
Combining this with (\ref{PASES}), gives that 
\be
R_t \ = \ r_t 
\ = \ \frac{\Es(K-1)}{n-1}
\ = \ 
R_{\mbox{ses}},
\ \ \ \ \ \ \ \ \
\mbox{for $t$ in the S.E.S.}.
\label{RrKRses}
\ee


\

We look now into how the generalized Hamilton's rule (\ref{HR+}) can be written in the  
case in which $v^A_k$ is a linear function of $k$. 
There is no loss in
generality in supposing that this linear function can be written as 
$v^A_k = -C + (k-1)B/(n-1)$, with $B,C$ constant.
We could, for instance, be considering the public
goods game, Example 1, or an additive pairwise interaction,
Example 4. But note that we are not making any assumption about $v^N_k$,
so that all that we are assuming is that at a cost $C$ to itself,
each altruist provides 
a benefit $B/(n-1)$ to each other altruist in its group. They may be
providing benefits to non-altruists or not, and if they do, the amount 
of that benefit is irrelevant for the current computation. 
We do not need, in particular, to assume linearity of $v^N_k$ in $k$.
We are also not yet assuming any restrictions on the values of $C$ and $B$.

The linearity of $v^A_k$ can be exploited 
by using 
(\ref{RrKRses})
to write
$$
\Es(v^A_K) \  =  \ -C + \frac{B}{n-1} \Es(K-1)  \ = \ 
-C + B R_{\mbox{ses}}.
$$
This transforms (\ref{HR+}) into the familiar
expression of Hamilton's rule:
\be
C \ < \ B \, \Rs,
\label{HRpg}
\ee
whenever $v^A_k = -C + (k-1)B/(n-1)$. 
But note that even in this case, in which $B$ and $C$ are 
constants, the relatedness 
$\Rs$ complicates the application of the rule. 
To compute $\Rs$ theoretically, one still needs information about $\nu$. 
And if analyzing data from an experiment that is supposed to be well 
modeled by our framework, one would have to sample from the S.E.S., 
not from the very early stage (before stationarity settles in), 
or from the late stage when the number of altruists is no longer negligible
as compared to the size of the whole population, so that the group type 
distribution is no longer directly related to $\nu$. 
For field data, this remark may make (\ref{HRpg}) of little value. 
Fortunately, this problem 
disappears in the case of weak selection, as we will see later.

Condition (\ref{HRpg}) refers to the viability 
of a single new mutant A. 
It is clear that if we started from any number of copies of the gene A 
that were negligible as compared to $g$, the same arguments would apply
and lead to the same rule. 
But had we started in generation 0 from 
a distribution of groups with a non-negligible fraction of altruists, 
(\ref{HRpg}) would not in general provide us with the direction of evolution.
Indeed, even the equilibria between alleles A and N, in the 
late stage of the evolution, will not in general turn the inequalities in 
(\ref{HRpg}) into equalities, unless $v_N = k B / (n-1)$. 
This point, that is emphasized in \cite{Vee1}, 
is well illustrated by considering
Example 4, in which $v^N_k = k B'/(n-1)$. As is well known, for arbitrary
$v^A_k$ and $v^N_k$, we can write the Price equation (\ref{price}), or (\ref{price++})
also in the following form, where $f = f(t)$:
$$
W(f) \, \E (p(t+1)  -  p(t) | f) \ = \ p(t)\, (1-p(t)) \, ( W^A(f) - W^N(f)),
$$
where
$$
W^A(f) \ = \ \frac{\sum_k w^A_k k f_k}{\sum_k k f_k} \ = \  \E_t(w^A_K|A_1),
$$
and
$$
W^N(f) \ = \ \frac{\sum_k w^N_k (n-k) f_k}{\sum_k (n-k) f_k} \ = \  \E_t(w^N_K|A^c_1).
$$
(The second inequality in each one of these last two displays is analogous to 
(\ref{PASES}), corresponding to averages over size biased samplings of the distribution $f$.)
In case $v^A_k = -C + (k-1) B /(n-1)$, and $v^N_k = k B'/(n-1)$, 
we can use (\ref{KPAA}) and (\ref{KPAAc}),
to reduce these to 
\begin{eqnarray*}
W^A(f) \ & = & \ 1 \ + \ \delta \, \E_t(-C + (K-1) B /(n-1)|A_1) \ = \ 1 \ + \ \delta \, (-C + B \P_t(A_2|A_1)),
\\ 
W^N(f) \ & = & \ 1 \ + \ \delta \, \E_t( K B' /(n-1)|A^c_1) \ = \ 1 \ + \ \delta \, B' \P_t(A_2|A^c_1).
\end{eqnarray*}
Hence, 
\begin{eqnarray*}
W^A(f) - W^N(f) 
\ &  = &   \
\delta \, (-C \ + \ B \P_t(A_2|A_1) \ - \ B' \P_t(A_2|A^c_1)) \\
\ &  = &   \
\delta \, (- C \ + \ B (\P_t(A_2|A_1) \ - \ \P_t(A_2|A^c_1)) \ 
+ \ (B - B') \P_t(A_2|A^c_1))  \\
\ &  = &   \
\delta \, (-C \ + \ B \, r_t \ + \ D \, (1-r_t) \, p(t)),
\end{eqnarray*}
where $D = B-B'$ is the synergy parameter, and we used 
(\ref{PAA}) and (\ref{rP-P}). 
When $0<p(t)<1$, and $\delta > 0$, 
the necessary and sufficient condition for \ $\E (p(t+1)|f(t)) \, > \, p(t)$, 
\ is therefore
\be
C \   <  \  B \, r_t \ + \ D \, (1-r_t) \, p(t).
\label{QR}
\ee
Condition (\ref{QR}) was derived in the case $n=2$ in 
\cite{Vee2}, who called it 
Queller's rule,
giving credit to the work in \cite{Que1}.
We refer the reader to \cite{SDZ}, for an extension of Queller's rule when $v^A_k$ and $v^N_k$ are 
not linear functions of $k$. 

\

Consider now the threshold model in Example 3.
In this case 
$\Es(v^A_K) = -C + A \Ps(K \geq \theta)$, 
so that the generalized Hamilton
rule (\ref{HR+}) reads
\be
C \ < \ A \, \Ps(K \geq \theta).
\label{HRtm}
\ee

Comparing (\ref{HRpg}) with (\ref{HRtm}) is elucidating. Both make the heuristic 
power of (\ref{HR+}) apparent: in both cases the cost incurred by carrying the 
gene A must be compensated by the expected benefit that it brings to its
carriers. We will address in more detain in the next paragraph the fact that this 
expected value is in the particular distribution $\Ps$, related to $\nu$. For the 
moment we focus on the fact that while (\ref{HRpg}) is in the usual Hamilton's 
rule form, (\ref{HRtm}) is not. We do not see anything deep about this difference. 
The special form of (\ref{HRpg}), involving relatedness, is a feature of the 
linearity of $v^A_k$, that allows one to break the computation of an expectation
into a sum of correlations between the genotype of the focal and each one of its
group companions, taken one at a time. In the case of (\ref{HRtm}), the threshold
nature of $v^A_k$, does not allow for such a decomposition in any meaningful and simple way,
as far as we can see. Trying to rewrite $\Es (v^A_K)$ in terms of the relatedness 
parameter $\Rs$ in this case, does not seem a natural idea. But there isn't anything
really that fundamental about relatedness in (\ref{HR+}). What is important there
is the distribution of the random variable $K$, i.e., the values of $\Ps(K = k)$, 
$k = 1, ...,n$. For a threshold model, the particular feature of this distribution 
that is of relevance is the tail probability $\Ps(K \geq \theta)$. For a linear model the 
relevant aspect of this distribution is the expectation $\Es(K)$, that is naturally
expressed in terms of relatedness, since $\Es(K) = \Es(K-1) + 1 = (n-1) \Rs + 1$,
by (\ref{RrKRses}).
Trying to rewrite (\ref{HRtm}) in terms of relatedness
seems as unnatural to us as trying to rewrite (\ref{HRpg}) in terms of the 
tail probabilities $\Ps(K \geq k)$.
In this connection, we refer the reader also to the conceptual discussion 
on this model, in the case $n=3$, in \cite{Vee1} (where it was called ``stag hunt game''),
\cite{Mar}, \cite{Vee2} and \cite{GWW}.
Especially important is the fact that in \cite{GWW} both, (\ref{QR}) and (\ref{HRtm})
are written in the traditional Hamilton form $c < b r$. This is accomplished there at
the cost of defining $c$ and $b$ as appropriate regression coefficients, that depend 
not only on parameters in the payoff functions, but also on the distribution of 
genes A and N in the population (see, for instance, (12) and (13) in that paper). 
Conceptually, this raises the question of what 
$c$ and $b$ mean in Hamilton's rule. 
Computationally, we have found it easier to compute 
directly with (\ref{QR}), (\ref{HRtm}) and more generally with 
the generalized Hamilton rule (\ref{rho>1nu}), or equivalently, (\ref{HR+}), 
rather then with the methods from \cite{GWW}.
Their $c$ and $b$ vary in time, as the distribution of genes A and N changes. 
For instance, to rewrite our (\ref{HRtm}) in their Hamilton form $c < b r$, one has to consider 
the distributions of these genes in the S.E.S.. Therefore, one has to first find 
$\nu$ and then use it in the computations of $c$ and $b$ as regression coefficients.
But once $\nu$ is obtained, (\ref{HRtm}) is available with little additional computational
work, since $\Ps(K \geq \theta) = \sum_{k \geq \theta} k \nu_k / \sum_{k = 1, ..., n} k \nu_k$.
For this matter, obtaining $\rho$ directly, as Perron-Frobenius eigenvalue of $M(A+B)$, 
and using (\ref{rho>1nu}) is, off course, easier than using either 
(\ref{HRtm}), or the methods from \cite{GWW}.
This raises the question whether (\ref{HRtm}) is ever of computational value. 
The answer is positive, since 
in Section 5 we will see that (\ref{HR+}) and, in particular,  
(\ref{HRtm}) yield simple and elegant formulas, 
when $\delta$ is small, and $n$ is large
(see (\ref{msapproxex3}), (\ref{msapproxex3'}) and the related Fig. \ref{fig14}). 
 
\

The viability condition (\ref{HR+}) and its special cases
(\ref{HRpg}) and (\ref{HRtm}) are heuristically meaningful, but computationally 
and experimentally they may not be such an advance over (\ref{rho>1nu}). The 
distribution $\nu$ is built into $\Ps (\cdot) $, and in particular into 
$\Rs$ and $\Ps (K \geq \theta)$. Computationally, finding $\nu$ is at least as 
demanding as finding $\rho$. Experimentally, one would not expect to sample in nature 
from the S.E.S., but rather from the late stage, after A has invaded and is now 
either fixated, or in a polymorphic equilibrium with N. (One can conceive, though,
lab experiments in which one could sample from the S.E.S..) Fortunately, these 
problems disappear when selection is weak, as we will see next.

\section{Weak selection and further conceptual discussion}
\label{sec4}

We turn now to simplifications to our analysis in case of small $\delta$, i.e., 
weak selection. In this case it is well known (see, e.g., 
\cite{Rou}, \cite{LKWR}) that there is a 
separation of time scales. Most of the time in most places evolution is occurring
as if $\delta$ were 0, i.e., via neutral genetic drift. Only occasionally
events happen that are caused by the slight differences in fitness of the individuals.
Mathematically, what one gains is the possibility of studying the process as a 
perturbation of the neutral evolution. For us, this will be encapsulated in a result that
we state next. We will include $\delta$ now as a superscript in the notation of 
quantities that depend on it (e.g., $M^{\delta}$, $\nu^{\delta}$, $\P^{A,\delta}_{\mbox{ses}}$, 
$R_{\mbox{ses}}^{\delta}$, $R^{\delta}_t$, etc).

We observe that when $\delta > 0$ is small enough, the generalized Hamilton rule (\ref{rho>1nu}) and its equivalent (\ref{HR+}) 
can be replaced by the requirement that
\be
\frac{\sum_k v^A_k k \nu_k^0}{\sum_k k \nu_k^0} 
\ = \
\sum_k \, v^A_k \, \Psz (K=k) \ = \ \Esz (v^A_K ) \ > \ 0.
\label{HR+ws}
\ee
Indeed, (\ref{HR+ws}) is an immediate consequence of (\ref{HR+}), due to the 
continuity of the Perron-Frobenius left-eigenvector $\nu^{\delta}$ of the 
matrix $M^{\delta}(A+B)$ as a function of $\delta$. 
As $\delta \to 0$, the vector $\nu^{\delta}$ converges to $\nu^{0}$, the 
Perron-Frobenius left-eigenvector of the matrix $M^0(A+B)$. Importantly,
$M^0(A+B)$, and hence also $\nu^0$, does not depend at all on the payoff 
functions. They depends only on $n$ and $m$. 
This is illustrated in Fig. \ref{fig8}, where one can see that $\nu^{\delta}$
is model dependent for large $\delta$, but becomes model independent under weak 
selection.  

The important simplification in (\ref{HR+ws}), with respect to (\ref{HR+}), is that 
$\nu^{\delta}$ has been replaced by $\nu^0$ in the left hand side of (\ref{HR+}). 
In particular, (\ref{HR+ws}) is not affected by the values of $v^N_k$. This 
may seem surprising at first sight, but can be understood as follows. 
In the E.S., most type N individuals are in groups without altruists, so that 
the mean fitness of the gene N is close to 1, not depending on the $v^N_k$. 
The values of the $v^N_k$ do affect the distribution $\nu^{\delta}$, because 
they affect the fate of the groups with altruists, and in this way affect the 
mean fitness of the gene A. (This is how the $v^N_k$ 
affect (\ref{HR+}) and its special cases (\ref{HRpg}) and (\ref{HRtm}).) 
But when $\delta$ is small, $\nu^{\delta}$ is close to $\nu^0$, 
and its dependence on $v^A_k$ and $v^N_k$ is a perturbation of order $\delta$. 
This dependence produces only a second order effect (of order $\delta^2$) on $\rho^{\delta}$.

Conceptually, one can see (\ref{HR+ws}) as a separation of the effects from
demographics from those of fitness. The distribution $\nu^0_k$ is
completely determined by the demographics, while the payoffs
$v^A_k$ carry only information about the fitness function. This
separation makes the concept of neighbor modulated fitness particularly
appealing in the regime of weak selection.
The viability condition (\ref{HR+ws}) says that the mutant gene
A is viable in case its average (neighbor modulated) fitness is
larger than that of the wild type N before A appeared, with the
weights in the average being given by the fashion in which the demographics 
alone arranges the genes in the population.

For theoretical analysis, (\ref{HR+ws}) is much simpler than (\ref{rho>1nu}), or (\ref{HR+}).
One can compute once and for all the distributions $\nu^0$, that depend only on 
$n$ and $m$. Then, in analyzing a model (given by $v^A_k$ and $v^N_k$), one simply 
looks at the average value of $v^A_k$ with respect to the universal distribution
$\nu^0$. This allows for a much greater intuition of what to expect, when comparing
models, than was possible in the case of strong selection, in which $v^A_k$ and $v^N_k$
also affect the distribution $\nu^{\delta}$. For instance, if two models I and II are
comparable in the sense that $v^{A,\mbox{I}}_k \leq v^{A,\mbox{II}}_k$, $k=1,...,n$, then we 
have $m_s^{\mbox{I}} \leq m_s^{\mbox{II}}$. This allows one to use a simple model,
like the threshold model in Example 3, to obtain estimates (say, lower bounds) on 
the value of $m_s$ for more interesting and realistic models, like those in 
Examples 2, 5 and 6.

The viability condition (\ref{HR+ws}) still refers to the S.E.S., but this can 
be overcome by considering a different random variable instead of $K$. 
Define $K^D$ as the number of individuals in the group to which the focal
belongs, that are IBD to the focal (the focal included in the counting). 
Clearly $K^D \leq K$,  
and, with overwhelming probability,  
$K^D = K$ in the E.S.. 
Observe that $\P_t^0(K^D = k | A_1) = \P_t^0 (K^D = k )$, for all $t$, since 
conditioning on $A_1$ does not affect lineages, when $\delta = 0$.  
But while $\P_t^0 (K = k | A_1 )$ depends on the fraction of altruists in the 
population, and so changes with time, $\P_t^0 (K^D = k )$ becomes constant 
for $t >> 1$ (or, more precisely, for $t >> 1/m$, so that by time $t$ 
it is likely that migration events have occurred in the focal's lineage).
We denote by $\pi = (\pi_1, ..., \pi_n)$ this equilibrium distribution:
$$
\pi_k  \ = \ \P^0_t (K^D = k), \ \ \ \ \ \mbox{for $t >> 1$}.
$$
Considering $t$ in the S.E.S., which has both $t >> 1$, and $K^D = K$, gives then
\be
\pi_k \ = \ \P^0_t (K^D = k) \ = \ \P^0_t (K^D = k | A_1) \ = \ \Psz (K = k) 
\ = \ \frac{k \nu^0_k}{ \sum_{k'} k' \nu^0_{k'}}.
\label{pinu}
\ee
And the viability condition (\ref{HR+ws}) can be rewritten as
\be
\sum_k \, v^A_k \, \pi_k \ > \ 0.
\label{HR+wspi}
\ee
This is a major improvement over (\ref{rho>1nu}) and (\ref{HR+}), 
since now no reference to the S.E.S. is left. 
The distribution $\pi$ only depends on the structure of the population and how genes
flow under neutral drift. Experimentally it can be accessed by sampling any neutral 
genetic markers from the population in demographic equilibrium (the condition $t >> 1$).

The distribution $\pi$ is directly associated to the relatedness $R^0_t$, $t >> 1$, in the 
same way that the distribution $\P_{\mbox{ses}}^{A,\delta} (K = k)$ is associated to 
$R_{\mbox{ses}}^{\delta}$.
Recall the enumeration of the members of the focal group, in which the first individual is 
the focal. Now 
decompose $K^D-1 = I^D_2 + ... + I^D_{n}$,
where, $I^D_j = 1$, if the $j$th individual
in this ordering is IBD to the focal, and $I^D_j = 0$, otherwise.
This yields $\E^0_t(K^D-1) = \E^0_t(I^D_2)+...+\E^0_t(I^D_{n}) = (n-1) \P^0_t(D)$,
and therefore
\be
R^0_t \ = \ \P^{0}_t (D|A_1) \ = \ \P^{0}_t (D) \ = \
\frac{\E^{0}_t (K^D) -1} {n-1} \ = \ 
\frac{\sum_k k \pi_k \,  - \, 1} {n-1}
 \  = \ 
R^0,
\label{Rpi}
\ee
for $t >> 1$, 
where the last equality introduces a new notation.

One can compute $R^0$ relatively easily, as follows.
If either the focal or the co-focal are
migrants, they are not IBD. If they are both non-migrants, they chose their parents
from the parental group of their group. With probability $1/n$ they chose the same
parent, and are therefore IBD. With probability $1-1/n$ they chose different parents.
In this last case they are IBD exactly if their parents are IBD, and this event has
probability $R^0$. Assembling these pieces, we obtain
$ R^0 \ = \ (1-m)^2 (1/n + (1 - 1/n)R^0)$ and hence,
\be
R^0 \ = \
\frac{(1-m)^2}{n-(n-1)(1-m)^2}
\ \approx \ \frac{1}{1+2nm},
\label{rm}
\ee
where the approximation is good when $m$ is small. This a well known
result by Wright for the infinite islands model, for haploid individuals.
It is important to clarify that even when $\delta = 0$, our framework is 
not identical to the infinite islands model. In that model there are 
a large number $g$ (to be taken to $\infty$ in the computations) of islands,
each one with $n$ individuals. In generation $t+1$ the $n$ individuals 
in each island choose, independently, a parent from the individuals in the 
same island in generation $t$. This is followed by migration at rate $m$, that 
is implemented in exactly the same way as in our framework. Differently from 
our framework, each island is ``parent'' to exactly one island in the next 
generation. In our framework, when $\delta = 0$, each group is parent to a random number, 
with distribution $\mbox{Bin}(g,1/g)$, of groups. In spite of this difference, it 
is not simply a coincidence that lead to the same formula (\ref{rm}) in both 
frameworks. They share the same coalescence structure, when we ask ourselves 
questions related to IBD. Not only is $R^0$ identical in these frameworks, 
but so is also the distribution $\pi$. Indeed, when we follow 
lineages backwards in time, until there are migration events, it does not 
matter if the group that we are following from generation to generation is
the same (as in the infinite islands case) or changes (as in our case). 

We will show next that 
\be
r^0_t = R^0,
\label{rR}
\ee
when $t >> 1$,
so that $R^0$ can also be obtained experimentally from standard statistical regression methods
(for instance using (\ref{FSTr})),
applied to neutral genetic markers, by sampling from a single generation $t$, without 
information about past generations or any knowledge about kinship.

To emphasize where in the proof of (\ref{rR}) the assumption $\delta = 0$ will matter, 
we first write, for arbitrary $\delta \geq 0$, 

\begin{eqnarray*}
r^{\delta}_t \  & = & \ \P^{\delta}_t (A_2|A_1) \ - \ \P^{\delta}_t (A_2|A_1^c)   \\
\   & = & \  
\P^{\delta}_t (A_2|D A_1) \, \P^{\delta}_t (D|A_1)
\ + \ \
\P^{\delta}_t (A_2|D^c A_1) \, \P^{\delta}_t (D^c|A_1)  \\
&  & -  \ \P^{\delta}_t (A_2|D A_1^c) \, \P^{\delta}_t (D|A_1^c)
\ - \
\P^{\delta}_t (A_2|D^c A_1^c) \, \P^{\delta}_t (D^c|A_1^c).  
\end{eqnarray*}
Clearly $\P^{\delta}_t (A_2|D A_1) = 1$, $\P^{\delta}_t (A_2|D A_1^c) = 0$.
The fact that $\delta = 0$ allows the following simplifications:
$$
R^0_t \ = \ \P^0_t (D|A_1) \ = \ P^0_t (D) \ = \ \P^0_t (D|A_1^c),
$$
so that also 
$$
\P^0_t (D^c|A_1) \ = \ 1 - R^0_t \ = \ P^0_t (D^c|A_1^c),
$$
and 
$$
\P^{0}_t (A_2|D^c A_1) \ = \ \P^{0}_t (A_2|D^c)  \ = \ \P^{0}_t (A_2|D^c A_1^c).
$$
(This would generally not hold with $\delta > 0$, since then information on the 
occurrence or not of $A_1$ biases the lineage of the co-focal, even if it does not 
meet that of the focal.) With the simplifications above, we readily obtain (\ref{rR}).

\

The identity (\ref{rm}) shows that, as expected, $R^0$ is a strictly decreasing function 
of $m$. This identity can be inverted as 
\be
m \ = \ 1 \ - \ \sqrt{\frac{n R^0}{(n-1) R^0 + 1}}.
\label{mr}
\ee
With fixed $n$ there is a 1-to-1 correspondence between $0 \leq m \leq 1$ and $0 \leq R^0 \leq 1$,
given by (\ref{rm}) and (\ref{mr}). It is natural then, in the regime of weak selection, 
to define the critical value of $R^0$ as 
\be
R^0_s \ = \
\frac{(1-m_s)^2}{n-(n-1)(1-m_s)^2}.
\label{R0s=R0s(ms)}
\ee
This is the least level of relatedness that makes the gene A viable in our framework, 
under weak selection.
Even when selection is not weak, we can define $R^0_s$ via (\ref{R0s=R0s(ms)}), 
since $R^0$ is a natural alternative way to identify the value of $m$, through
(\ref{rm}), and can be measured using neutral genetic markers. 
See Fig. \ref{fig13} for the shape, on a logarithmic scale of the function (\ref{R0s=R0s(ms)}).

\

Under weak selection, 
when $v^A_k = -C + (k-1) B / (n-1)$ is a linear function of $k$, as in the public goods game,
or additive pairwise interactions,
the viability condition (\ref{HRpg})
now reads
\be
C \ < \ B \, R^0.
\label{HRpgws}
\ee
This is a fully standard Hamilton rule, with $R^0$ independent of $B$ and $C$, being 
a function only of the population structure, through $n$ and $m$. 
The critical values are therefore
\be
R^0_s \ = \ C/B, \ \ \ \ \ \
\mbox{or, equivalently,} 
\ \ \ \ \ \ 
m_s \ = \ 1 - \sqrt{\frac{Cn}{B + C(n-1)}} \ \approx \ \frac{B-C}{2Cn},
\label{R0sforEX1}
\ee
where we used (\ref{mr}), and
the approximate result is valid when $n$ is large.


Condition (\ref{HRpgws}) is an expression in our setting of the common wisdom 
according to which Hamilton's rule in its standard form applies when selection 
is weak and interactions are pairwise and additive (Example 4, in our setting).
In this regard, there are several points worth commenting.

First, (\ref{HRpgws}) holds with no
assumption on $v^N_k$, so that it goes beyond that common wisdom. 

Second, that as explained after (\ref{HRpg}), we are only considering a viability 
condition for an initial situation with a single (or for this matter, a number 
$<< g$) of type A individuals. Even in the case of weak selection, and with pairwise
additive interactions (so that $v^A_k = -C + (k-1) B/(n-1)$, $v^N_k = k B' / (n-1)$), 
when the numbers of type A and type N are comparable to $g$, the condition for
selection to increase the frequency of type A is given by Queller's rule 
(\ref{QR}), that now reads

\be
C \  < \ B \, R^0 \ + \ D \, (1-R^0) \, p,
\label{QRws}
\ee
where, as before, $D = B' - B$ and $p$ is the fraction of type A in the population. 

Third, the assumption of weak selection greatly simplifies relatedness, 
and makes it more universal, but otherwise, also under strong selection (\ref{HRpg})
has the usual Hamilton rule form.

Fourth, in our setting it is not the pairwise additive nature of the interaction that 
matters. What matters for (\ref{HRpg}) and (\ref{HRpgws})
to hold is linearity of $v^A_k$ in $k$ (and for (\ref{QR}) and (\ref{QRws}) to also hold,
additional linearity of $v^N_k$ in $k$), so that one can use (\ref{KPAA}) and (\ref{KPAAc}). 
In Example 4, we explained that this sort of linearity 
is formally equivalent to having pairwise additive interactions, but it can also
result from interactions involving many individuals at the same time, as in Example 1.

Fifth, the meaning of $C$ and $B$ in (\ref{HRpgws}) has to be carefully understood. 
This point is well illustrated by considering the fitnesses in (\ref{ipd}), that corresponds to 
the case of pairwise interactions in iterated prisoners dilemmas, with types N always defecting 
and types A playing tit-for-tat. In this case $C =c  (n-1)$ and 
$B = ((b-c)T+c)(n-1)$, where $c$ is the cost to an actor each time it cooperates, $b$
is the benefit that an actor provides to its partner each time it cooperates, and $T$ 
is the average number of iterations of each pairwise interaction. 
The viability condition (\ref{HRpgws}) now reads $c < ((b-c)T+c) R^0$ and we have 
\be
R^0_s \ = \ \frac{c}{(b-c)\, T \, + \, c}.
\label{ipgR0s}
\ee
For $T = 1$ we have $R^0_s \, = \, c/b$, but $R^0_s$ decreases as $T$ increases. 
Nothing here is surprising, but these computations illustrate the fact that 
$C$ and $B$ are life-cycle costs and benefits, and not costs and benefits in 
each momentary interaction (even if those are constant, as in the present case). 

\

In contrast to (\ref{HRpgws}), the weak selection form of the viability 
condition (\ref{HRtm}), for the threshold model, Example 3, 
reads
\be
C \ < \ A \,\sum_{k \geq \theta} \pi_k.
\label{HRtmws}
\ee
The quantity $\sum_{k \geq \theta} \pi_k$ 
depends only on $n$ and $m$. 
Thanks to (\ref{mr}), it can also be seen as a function of $n$ and $R^0$ only.
But we see no hope in expressing this functional dependence in simple terms, 
that would allow us to derive a simple expression for $R^0_s$ in this example.
We will, nevertheless, obtain interesting approximations for it in Section 5.
We will also be able there to compute the exact limit of $R^0_s$ for this example, 
as $n \to \infty$, provided that $a = \widetilde a \, n$ for some constant 
$0 \leq \widetilde a \leq 1$. 


\

To apply the viability condition (\ref{HR+wspi}), we need to compute $\pi$.
One solution is to find $\nu^0$, and use (\ref{pinu}). 
And $\nu^0$ can be computed as the Perron-Frobenius left-eigenvector of
$M^0(A+B)$, where 
$$
(M^0)_{k,k'} \ = \ \P \, (\mbox{Bin} (n, k/n) = k').
$$
This yields, after some simplifications, 
\be
\left( M^0(A+B)\right)_{k,j}  
\ = \
\left \{
\begin{array}{ll}
                    \P\,(\mbox{Bin} (n, (1-m)k/n) = j) \ + \   
                     m k, & \mbox{if \  $j = 1$,} \\
                     \P \, (\mbox{Bin} (n, (1-m)k/n) = j),    & \mbox{if \ $j = 2, ..., n$.}
\end{array}
\right.
\label{M0AB}
\ee
Moreover, one can use the fact that $\rho^0 = 1$ to further simplify the computation of $\nu^0$. 

\begin{figure}[th!]
\begin{center}
\includegraphics[width=6cm]{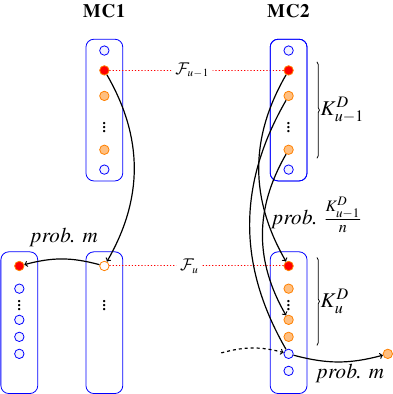}
\end{center}
\caption{This diagram illustrates the discussion that leads to (\ref{Q}). $K^D_{u}$ represents the number 
of individuals that are IBD to a focal individual ${\cal F}_{u}$ (red circle) in generation $u$. Two scenarios are 
discernible for the previous generation $u-1$. MC1 (left panel): the focal individual is a migrant. This can happen 
with probability $m$ and, in the ES for $g\rightarrow\infty$, implies that $K^D_{u}=1$. MC2 (right panel): the focal 
individual  ${\cal F}_{u}$ (red circle) is a child of ${\cal F}_{u-1}$.  Each individual in the focal group in generation 
$u$ choses a parent from the group of ${\cal F}_{u-1}$ in the  previous generation with uniform probability, as $\delta=0$. 
With probability $K^D_{u-1}/n$ a parent is IBD with the focal individual  ${\cal F}_{u-1}$ (orange circles) 
and, consequently, his children are  also IBD with ${\cal F}_{u}$. Additionally, each individual in
 generation $u$ can  migrate with probability $m$. The number of IBD individuals in generation $u$ is, therefore, 
 the focal individual himself plus a number of individuals  given by a binomial random variable  
with probability of success  $(1-m)K^D_{u-1}/n$ in $n-1$ trials. }
\label{fig10}
\end{figure}

One can also use, alternatively, methods from coalescence theory, 
to study the distribution $\pi$.
But because we find these methods quite cumbersome for this 
purpose, we introduce next a further alternative approach,
illustrated in Fig. \ref{fig10}, and explained next. 

Consider again the random experiment of choosing a focal individual from 
generation $t$. 
Denote by ${\cal F}_u$, $u=0,1,...,t$, the ancestor of the focal in generation $u$,
so that, in particular, ${\cal F}_t$ is the focal. Denote by ${\cal G}_u$ the 
group to which ${\cal F}_u$ belonged. 
Then denote by $K^D_u$ the number of members of ${\cal G}_u$ that where IBD to
the focal's ancestor ${\cal F}_u$ (including ${\cal F}_u$ itself). 
Our basic observation is that, when $\delta = 0$, so that there is no selection, 
the sequence of random variables $K^D_0,K^D_1, ..., K^D_t$ forms 
a time-stationary Markov chain on the set $\{1,...,n\}$,
described as follows.  Given the value
of $K^D_{u-1}$, the value of $K^D_u$ is:

\begin{description}
\item(MC1) With probability $m$, set to 1. [Migration event in focal's line of descent.]

\item(MC2) With probability $1-m$ set equal to
$1 + \mbox{Bin} (n-1, (1-m) K^D_{u-1}/n)$. [No migration event in focal's line of descent.]

\end{description}
This corresponds to a transition matrix $Q$ (that does also not depend on $t$) given by:
\be
Q_{i,j} \ = \ 
\left\{
                \begin{array}{ll}
m \ + \ (1-m) \, \P \, (\mbox{Bin} (n-1,(1-m)i/n) = 0),  & \mbox{if \  $j = 1$},  \\
(1-m) \, \P \, (\mbox{Bin} (n-1,(1-m)i/n) = j-1), & \mbox{if \  $j = 2,...,n$}.
                \end{array}
\right.
\label{Q}
\ee
To understand this claim, first observe that if the ancestor of the focal in 
generation $u$, ${\cal F}_u$ was a migrant, then, by definition of IBD, we 
have in ${\cal G}_u$ no one other than ${\cal F}_u$ itself IBD to ${\cal F}_u$, and hence
$K^D_u = 1$.
This corresponds to (MC1) above.
On the other hand, if ${\cal F}_u$ 
was not a migrant, then the number, $K^D_u - 1$ of other members of ${\cal G}_u$ that were 
IBD to ${\cal F}_u$
is easily obtained as follows. Each one of the $n-1$ other members
of ${\cal G}_u$ chose independently a parent from ${\cal G}_{u-1}$.
The group ${\cal G}_{u-1}$ had $K^D_{u-1}$ members that were IBD to
${\cal F}_{u-1}$ (with ${\cal F}_{u-1}$ included). 
Members of ${\cal G}_u$ other than ${\cal F}_u$ became IBD to ${\cal F}_u$
if two conditions were satisfied: they had to choose a member of 
${\cal G}_{u-1}$ that was IBD to ${\cal F}_{u-1}$, and they had to stay in
the group ${\cal G}_u$, rather than migrate out. (If they migrated out,
their replacements would not be IBD to ${\cal F}_u$, by definition of IBD).
Each one of these $n-1$ individuals in ${\cal G}_u$ had, therefore, independently,
probability $(1-m) K^D_{u-1}/n$ of being IBD to ${\cal F}_u$. 
This gives for $K^D_u -1$ the binomial probability in (MC2).
Adding 1, for ${\cal F}_u$ itself, gives us the full expression in (MC2).


The Markov chain $K^D_u$, starts from $K^D_0 = 1$, and when $u >> 1$ 
it will have reached its stationary state $\pi$.
(It is clear that the chain is irreducible and aperiodic when $0 < m < 1$, since 
in this case all entries of $Q$ are strictly positive. In case $m=0$, it is 
clear that the chain converges to its single absorbing state $n$, so that 
$\pi = (0,0, ..., 0, 1)$. In case $m = 1$, it is clear that 
the chain converges to its single absorbing state $1$, so that 
$\pi = (1,0,0,...,0)$.)
One can compute $\pi$ using stationarity,
by solving the linear system of equations 
\be
\pi \, Q \ = \ \pi, \ \ \ \ \ \ \ \sum_k \, \pi_k \ = \ 1.
\label{piQ}
\ee

Identity (\ref{pinu}), relating $\pi$ and $\nu^0$, can now be alternatively derived 
by observing that, from (\ref{M0AB}) and (\ref{Q}), we have 
$j (M^0(A+B))_{k,j} = k Q_{k,j}$.

\

The distribution $\pi$ enjoys a nice monotonicity property, as a function of $m$.
We recall that a probability distribution $\eta$ over $\{1,...,n\}$ is said to
be stochastically larger than another one, $\zeta$, if
$\sum_{k \geq k_0} \eta_k \ \geq \ \sum_{k \geq k_0} \zeta_k$, for $k_0 = 1,...,n$.
In this case we write $\eta \succeq \zeta$. It is known that this relationship is
equivalent to the statement that $\sum_{k}\eta_k h_k \ \geq \ \sum_{k}\zeta_k h_k$,
whenever $h_k$ is increasing in $k$.
We claim that $\pi$ is stochastically monotone decreasing in $m$, i.e.,
\be
\pi(m) \ \succeq \ \pi(m'), \ \ \ \ \ \mbox{whenever} \ \ \ \ \ m' \ \geq \  m.
\label{pimonotone}
\ee
Therefore,
under (C4), the left hand side of (\ref{HR+wspi}) is decreasing in $m$. In particular,
there can only be one value of $m_s$ that satisfies $\sum_k \, v^A_k \, \pi_k(m_s) \ = \ 0$.
The claim (\ref{pimonotone}) follows from the double observation that $\eta \, Q(m)$ is stochastically decreasing in
$m$ and stochastically 
increasing in $\eta$. (To see this more easily, consider the description of $Q$ in (MC1) and (MC2)
above.  Note that with $K^D_u$ fixed, increasing $m$ decreases $K^D_{u+1}$, and
that with $m$ fixed, $K^D_{u+1}$ is increasing with $K^D_u$.) We can use the monotonicity in
$m$ in this observation, to write, whenever $m' \ \geq \  m$,
$$
\pi(m) \ = \ \pi(m) \, Q(m) \ \succeq \ \pi(m) \, Q(m').
$$
Now, we can use the monotonicity in $\eta$ in the observation above to iterate this inequality
and obtain $\pi(m)\  \succeq \ \pi(m) \, (Q(m'))^t$, for arbitrary $t$. Letting $t \to \infty$,
gives $\pi(m) \ \succeq \ \pi(m')$.

\

The Markov chain introduced above 
can also be used in other ways,
by allowing one to write recursions for quantities of interest. 
For instance, this method can be used for computing the 
moments of the distribution $\pi$, \
${\cal M}_l = \sum_k k^l \pi_k$, \ $l=1,2,...$.
This is in principle
very useful, since an arbitrary $v^A_n$ can always be approximated by a polynomial.
So, in theory, one can compute $\E(v^A_K)$ in good approximation and, through
(\ref{HR+wspi}) compute $m_s$ also in good approximation. 
We will see in the next section, that in spite of the expressions for the moments 
${\cal M}_l$ being quite involved, they provide very powerful information. 
The heuristic nature of the recursive method requires going behind the apparently cryptic 
expression for $Q$, and instead using its description in items (MC1), (MC2), 
that appears immediately before $Q$ was introduced. 
We illustrate the method computing first the mean of $\pi$, \ ${\cal M} = {\cal M}_1$.
When the Markov chain 
is stationary, both $K^D_{u-1}$ and $K^D_u$ have distribution $\pi$. 
Therefore, from (MC1) and (MC2) we obtain:
$$
{\cal M} \ = \ m \ + \ (1-m) \left\{ 1 + (n-1) (1-m) {\cal M}/n \right\}.
$$
This yields
\be
\sum_k k \pi_k 
\ = \
{\cal M}
\ = \ 
\frac{n}{n-(n-1)(1-m)^2}.
\label{EK}
\ee
This result could also have been obtained by combining (\ref{Rpi}) and (\ref{rm}), 
or alternatively (\ref{rm}) could have been obtained from (\ref{Rpi}) and 
(\ref{EK}).

To express the $l$th moment, ${\cal M}_l$, in terms of ${\cal M}_j$, $j = 1,2,...,l-1$, 
we will use the following fact. If $X$ has a $\mbox{Bin}(N,p)$ distribution, then 
$$
\E(X^l) \ = \ \sum_{j=1}^l \, S(l,j)\, N_j \, p^j, 
$$
where $N_j = N(N-1) \cdots (N-j+1)$, and the Stirling number of the second kind, $S(l,j)$, is
the number of ways in which a set with $l$ elements can be partitioned into $j$ non-empty sets.
Clearly $S(l,1) = S(l,l) = 1$, for all $l$. They are also known to satisfy
$$
S(l,j) \ = \ \frac{1}{j!} \, \sum_{i=0}^j \left( \begin{array}{c} j \\ i \end{array} \right)
\, (j-i)^l.
$$
The first few values of $S(l,j)$ are: $S(1,1) = 1; S(2,1) = 1, S(2,2) = 1; S(3,1) = 1, S(3,2) = 3,
S(3,3) = 1; S(4,1) = 1, S(4,2) = 7, S(4,3) = 6, S(4,4) = 1; S(5,1) = 1, S(5,2) = 15, S(5,3) = 25,
S(5,4) = 10, S(5,5) = 1; ...$

For our purpose,
it is convenient to write $s = 1-m$, and use the Markov chain description (MC1), (MC2) as it
applies to $(K^D_u - 1)^l$: With probability $m$ this quantity takes the value 0, and with 
probability $s$ it takes the value $(\mbox{Bin}(n-1,sK^D_{u-1}/n))^l$. Therefore 
$$
\E((K^D_u - 1)^l) \ = \ s \, \sum_{j=1}^l \, S(l,j) \, (n-1)_j \, s^j \E(K^D_{u-1})^j / n^j.
$$
Since in equilibrium, $K^D_{u-1}$ and $K^D_u$ both have distribution $\pi$, we obtain 
$$
\sum_{j=0}^l \, \left( \begin{array}{c} l \\ j \end{array} \right) \, (-1)^{l-j} \, {\cal M}_j
\ = \ \sum_{j=1}^l \, S(l,j) \, \frac{(n-1)_j}{n^j} \, s^{j+1} \, {\cal M}_j.
$$
This provides us with the aimed recursion:
\be
{\cal M}_l \ = \ \frac
{
(-1)^{l+1} \ + \ \sum_{j=1}^{l-1} \, \left\{ 
(-1)^{l-j+1} \, \left( \begin{array}{c} l \\ j \end{array} \right) \ + \ S(l,j) \, \frac{(n-1)_j}{n^j} \, s^{j+1}
\right\} {\cal M}_j
}
{
1 \ - \ \frac{(n-1)_l}{n^l} \, s^{l+1}
}.
\label{calMl}
\ee

To illustrate the use of (\ref{calMl}), we insert the value of ${\cal M}_1 = {\cal M}$, from 
(\ref{EK}), into the recursion for ${\cal M}_2$:

\begin{eqnarray*}
{\cal M}_2 \ & = & \ 
\frac{ -1 \ + \ \left\{ 2 \ + \ \frac{n-1}{n} \, s^2 \right\} {\cal M}_1} 
    {1 \ - \ \frac{(n-1)(n-2)}{n^2} \, s^3 }
\ = \ 
\frac{ -n^2 \ + \ \left\{ 2n^2 \ + \ n(n-1)s^2 \right\} \, \frac{n}{n - (n-1)s^2}} 
    {n^2 \ - \ (n-1)(n-2) s^3 }
\\
\ & = & \ \frac{ n^2 \, (n + 2(n-1) s^2) }{(n^2 - (n-1)(n-2)s^3)\, (n-(n-1)s^2)}.
\end{eqnarray*}
This gives for the variance of the distribution $\pi$ the value:
$$
\sum_k \, (k \, - \, {\cal M})^2 \, \pi_k \ = \ {\cal M}_2 \ - \ ({\cal M}_1)^2
\ = \  
\frac{n^2 \, (n-1) \, s^2 \, (n  + (n-2) s - 2(n-1)s^2) }
{ (n^2 - (n-1)(n-2)s^3) \, (n - (n-1)s^2)^2}.
$$

\

The expressions above for ${\cal M}_l$ and the variance
are obviously very involved. 
They simplify substantially when $n$ is large, since $(n-1)_j/n^j \to 1$ 
as $n \to \infty$. 
This is not a surprise, since $n$ does not appear in (MC1), and in (MC2) the binomial
random variable converges in distribution to a Poisson random variable, with mean 
$s K_{u-1}^D$. The resulting Markov chain has state space $\{1,2,...\}$, but it is not 
hard to show that it is positive recurrent, and so has a single stationary distribution,
to which $\pi$ converges as $n \to \infty$. We will not explore this limit here, 
but rather study, in the next section, a limit in which as $n \to \infty$, also
$m \to 0$. As we will see, in this limit $\pi$ simplifies considerably, and leads 
to a number of very interesting applications.

\section{Limit of large {\it n} and small {\it m} under weak selection}
\label{sec5}


In this section we will continue to assume that selection is weak  
and we will study the limit in which
\be
\mbox{$n \to \infty$, \, $m \to 0$ \, and \, $n m \to \widetilde m$, \, so that \,
 $R^0 \to 1/(1 + 2 \widetilde m) = \widetilde R^0$},
\label{nmlimit}
\ee
where we used (\ref{rm}). To state the results on the behavior 
of the distribution $\pi$ in this limit, we suppose that $K^D$ is a
random variable with distribution $\pi$, i.e., $\P(K^D = k) = \pi_k$.
We state and comment the results in (a) and (b) below, and
afterwards, explain how to do the computations.
(See also Fig. \ref{fig12}.) 

\

\noindent (a) If $0 \leq \widetilde m < \infty$, then for $l = 1, 2, ...$, 
\be
\frac{{\cal M}_l}{n^l} \ \longrightarrow \ 
\frac{l!}{(2 \widetilde m + 1)(2 \widetilde m + 2) \cdots (2 \widetilde m + l)}
\ = \ 
\widetilde {\cal M}_l.
\label{KD/n}
\ee
This implies (see, e.g., Section 2.3.e of \cite{Dur})
that the random variable $K^D/n$ converges in distribution to 
a distribution with $l$th moment
$\widetilde {\cal M}_l$.

When $0 < \widetilde m < 1$,
these moments characterize a Beta distribution, with parameters $\alpha = 1$
and $\beta = 2 \widetilde m$, which has density
$f_{\widetilde m}(x) \ = \ 
2 \widetilde m \, (1-x)^{2 \widetilde m -1}$,
$0 <  x < 1$. In other words, 
\be
\P(K^D/n > x) \ \to \ 
\int_x^1 f_{\widetilde m}(x') dx' \ = \ 
\ (1-x)^{2 \widetilde m},
\label{KDntoBeta}
\ee
for arbitrary $0 \leq x \leq 1$. 
Notice that, for each $x$, this tail probability, $(1-x)^{2 \widetilde m}$, is decreasing in 
$\widetilde m$, meaning that the corresponding family of Beta distributions
is stochastically monotone decreasing in $\widetilde m$. 

The case $\widetilde m = 1/2$, which has $\widetilde R^0 = 1/2$, is particularly simple. 
In this case the limiting distribution, Beta(1,1) is the uniform distribution 
between 0 and 1, 
with $f_{\widetilde m}(x) = 1$, $0 < x<1$.
This case can be seen as separating two qualitatively distinct 
cases: When $0 < \widetilde m < 1/2$, the density $f_{\widetilde m}(x)$ is increasing 
in $0 < x<1$, while when $1/2 < \widetilde m < \infty$, the density $f_{\widetilde m}(x)$ is decreasing
in $0 < x<1$. In the extreme cases, in which $\widetilde m$ is close to 0 or very large, 
the density $f_{\widetilde m}(x)$ concentrates, respectively, close to 1 or 0. 

When $\widetilde m = 0$, we have $\widetilde {\cal M}_l = 1$, for each $l = 1, 2, ...$. 
These are the moments of a degenerate random variable, that takes the value 1 with probability
1. Therefore the random variable $K^D/n$ converges in probability to 1. 

\

\noindent (b) If $ \widetilde m = \infty$, we have 
for $l = 1, 2, ...$,
\be
\frac{{\cal M}_l}{n^l} \ \longrightarrow \ 0.
\label{KD/ninfty}
\ee
These are the moments of a degenerate random variable, that takes the value 0 with probability
1. Therefore the random variable $K^D/n$ converges in probability to 0.

But a different way of scaling $K^D$, provides a non-degenerate limit. We have, for $l = 1, 2, ...$,
\be 
m^l \, {\cal M}_l \ \longrightarrow \ \frac{l!}{2^l},
\label{KDm}
\ee
i.e., $m \, K^D$ converges in distribution to a random variable
with $l$th moment $l!/2^l$.
These moments characterize an exponential 
distribution with mean $1/2$. In other words, 
\be
\P(m K^D > x) \ \to \
\exp(-2x), 
\label{mKDexp}
\ee
for arbitrary $x \geq 0$.

\

The claims in (\ref{KD/n}), (\ref{KD/ninfty}) and (\ref{KDm}),
about the convergence of the  scaled moments, can be readily obtained by 
induction in $l = 1, 2, ...$, from (\ref{EK}) and (\ref{calMl}), and the following 
two observations. First, for $j = l-1$, 
\begin{eqnarray*}
(-1)^{l-j+1} \, \left( \begin{array}{c} l \\ j \end{array} \right) \ + \ S(l,j) \, \frac{(n-1)_j}{n^j} \, s^{j+1}
\ = \
\left( \begin{array}{c} l \\ l-1 \end{array} \right)
\ + \
\left( \begin{array}{c} l \\ 2 \end{array} \right)
\, \frac{(n-1)_{l-1}}{n^{l-1}} \,  s^l                        \\
\ \longrightarrow \
l \ + \ \frac{l\, (l-1)}{2}
\ = \
\frac{l\, (l+1)}{2}.
\end{eqnarray*}
Second, the denominator in (\ref{calMl}) can be rewritten as 
\begin{eqnarray*}
1 \ - \ \frac{(n-1)_l}{n^l} \, s^{l+1} 
\ & = & \ 
1 \ - \ 
\left( 1 - \frac{1}{n} \right) \, 
\left( 1 - \frac{2}{n} \right) \, ... \, 
\left( 1 - \frac{l}{n} \right) \,
(1 \, - \, m)^{l+1}
\\
\ & = & \ 
\frac{1+2+...+l}{n} \ + \ (l+1) \, m \ + \ \Delta(n,m) 
\\
\ & = & \ 
\frac{l \, (l+1)}{2n} \ + \ (l+1) \, m \ + \ \Delta(n,m),
\end{eqnarray*}
where $\Delta(n,m)/m \, \to \, 0$ and $\Delta(n,m) \, n \, \to \, 0$.

\



A question that comes naturally to mind is whether for the threshold model in Example 3, 
we can find the exact value of $m_s$, in the case of weak selection. Unfortunately we have not
been able so far to 
compute the exact value of $\sum_{k \geq \theta} \pi_k$.
We can, nevertheless, use the result in (\ref{KDntoBeta}) above, that states that when 
$n$ is large and $m$ is small, then 
$\sum_{k \geq \theta} \pi_k \ \approx \ (1-\theta/n)^{2mn}$. 
Using this approximation in combination
with (\ref{HRtmws}), yields
\be
m_s \ \approx \ \frac{ \log(C/A)} {2n \, \log(1-\theta/n)}.
\label{msapproxex3}
\ee
This approximation should be good when 
$n$ is large, and $\theta/n$ is not too close to 0, so that the resulting value of $m_s$ 
is small.
When, additionally, $\theta/n$ is substantially smaller than 1, we can further approximate
\be
m_s \ \approx \ \frac{1}{2\theta} \, \log \left( \frac{A}{C} \right).
\label{msapproxex3'}
\ee
See Fig. \ref{fig14}, for a comparison of the exact value of $m_s$ under weak selection, from 
the viability condition (\ref{HR+wspi}) (or, equivalently, (\ref{HRtmws}))
and the approximations (\ref {msapproxex3}) 
and (\ref{msapproxex3'}). Surprisingly, (\ref {msapproxex3}) gives a good approximation
there even when $\theta$ is small.  

\ 

The approximation (\ref{msapproxex3}) for Example 3 can be formalized and extended to 
a wide class of models in the following way. 
(See Fig. \ref{fig16}, for an illustration.)
Suppose that there is a piecewise continuous
and bounded function $\widetilde v^A_x$, \, $0 \leq x \leq 1$, 
such that 
\be
\max_{k = 1, ...,n} 
\ \left| v^A_k - \widetilde v^A_{k/n} \right| 
\ \longrightarrow \ 0, \ \ \ \ \ \mbox{as \  $n \to \infty$}.
\label{limv}
\ee
Then, the result in (\ref{KDntoBeta}) above, combined with the continuous mapping theorem, (2.3) in Section 
2.2.b of \cite{Dur}, implies that, when $0 < \widetilde m < \infty$,  
\be
\sum_k \, v^A_k \pi_k \ \longrightarrow \ \int_0^1 \widetilde v^A_x \, f_{\widetilde m}(x) \, dx
\ = \ 2 \widetilde m \, \int_0^1 \widetilde v^A_x \, (1-x)^{2\widetilde m -1} \, dx 
\ = \ \widetilde V^A(\widetilde m),
\label{wVA}
\ee
in the limit (\ref{nmlimit}),
where the last equality introduces a new notation.
This means that, in this limit, 
the viability condition (\ref{HR+wspi})
now reads
\be
\widetilde V^A(\widetilde m) \ > \ 0, \ \ \ \ \ \mbox{or, equivalently,} \ \ \ \ \ 
\int_0^1 \widetilde v^A_x \, (1-x)^{2\widetilde m -1} \, dx \ > \ 0.
\label{vcnmlimit}
\ee
We want to define $\widetilde m_s$ as the solution of
\be
\widetilde V^A(\widetilde m_s) \ = \ 0.
\label{wms}
\ee
To assure that this equation has a solution it is sufficient to assume that 
$\widetilde v^A_x$ is continuous at the end-points $x = 0$ and $x =1$. 
In this case, under under (C1),
$$
\widetilde V^A(\widetilde m) \ \to \ \widetilde v^A_0 \ = \ v^A_1 \ < \ 0,
$$
as $\widetilde m \to \infty$.
And under (C2),
$$
\widetilde V^A(\widetilde m) \ \to \ \widetilde v^A_1 \ = \ v^A_n \ > \ 0,
$$
as $\widetilde m \to \ 0$.
Since clearly $\widetilde V^A(\widetilde m)$ is continuous in $\widetilde m$, (\ref{wms}) then 
must have a solution  $0 < \widetilde m_s < \infty$. If there is more than one solution, 
then we define $\widetilde m_s$ as the largest one. 
If condition (C4) holds, then 
$\widetilde v^A_x$ is non-decreasing in $0 < x < 1$. 
In this case, the stochastic monotonicity of the Beta distributions in (a) imply that 
$\widetilde V^A(\widetilde m)$ is non-increasing in $\widetilde m$. 
Actually, it is clear, from the behavior of the density 
$f_{\widetilde m}(x) \, = \, 2\widetilde m \, (1-x)^{2\widetilde m -1}$,
that $\widetilde V^A(\widetilde m)$ is then strictly increasing 
in $\widetilde m$, unless $\widetilde v^A_x$ is constant. 
Hence, when 
$\widetilde v^A_x$ is continuous at 0 and 1 and 
(C1), (C2) and (C4) hold, (\ref{wms}) defines $\widetilde m_s$ uniquely.

The viability condition (\ref{vcnmlimit}) implies 
that if we consider a 
model that satisfies (\ref{limv}), then in the weak selection regime, 
\be
n \, m_s  \, \to \, \widetilde m_s \ \ \ \ \ \
\mbox{and} \ \ \ \ \ \
R^0_s \, \to \, \widetilde R^0_s \, = \, \frac{1}{1+2\widetilde m_s}, 
\ \ \ \ \ \ \ \ 
\mbox{as} \ \ \ \ \ n \, \to \, \infty.
\label{nmstowms}
\ee

\

Example 3 satisfies (\ref{limv}), if we take 
$\theta \, = \, \lceil \, n \, \widetilde \theta \, \rceil$, 
for some $0 \leq \widetilde \theta \leq 1$, 
where $\lceil y \rceil$ is the smallest integer larger than or equal to $y$.
In this case, $\widetilde v^A_x = -C$, for $0 \leq x < \widetilde \theta$,
and $\widetilde v^A_x = -C + A$, for $\widetilde \theta \leq x \leq 1$.
This yields $V^A(\widetilde m) \, = \, -C \, + \, A\, (1-\widetilde \theta)^{2 \widetilde m}$,
and therefore 
$$
\widetilde m_s \ = \ \frac{ \log(C/A)}{2 \, \log(1-\widetilde \theta)},
\ \ \ \ \ 
\mbox{or, equivalently,}
\ \ \ \ \ 
\widetilde R^0_s \ = \ \frac{1}{1 +
\frac{ \log(C/A)}{\log(1-\widetilde \theta)} }.
$$  
The approximation (\ref{msapproxex3}) can be seen now as a special case of (\ref{nmstowms}). 

\

Example 2 also satisfies (\ref{limv}), if we take 
$a \, = \, \lfloor  n \, \widetilde a \rfloor$, for some 
$0 \leq \widetilde a \leq 1$, where $\lfloor y \rfloor$ is the 
largest integer smaller than or equal to $y$ (the integer part of $y$).
In this case, $\widetilde v^A_x \, = \, -C + Bx$, for $0 \leq x \leq \widetilde a$,
and $\widetilde v^A_x \, = \, T \, (-C + Bx)$, for $\widetilde a < x \leq 1$.
(See Fig. \ref{fig16}.)
Condition (\ref{mtft}) is satisfied, when $n$ is large, in case 
$\widetilde a > C/B$, and fails if $\widetilde a < C/B$.
In case $\widetilde a = C/B$, (\ref{mtft})
may be satisfied or fail, depending on whether $n \widetilde a$ is close to $a$ or $a+1$,
but in either case, the left hand side of (\ref{mtft}) is 
of order $1/n$, so that it is only 
marginally satisfied or violated.
In our analysis below, we will not assume that (\ref{mtft}) holds.
We have, for this model, 
$$
V^A(\widetilde m) \ = \ \frac{B}{2\widetilde m + 1} \ - C \ + \ 
(T-1) \, (B \widetilde a - C)\,(1-\widetilde a)^{2\widetilde m} \ + \
(T-1) \, \frac{B \, (1-\widetilde a)^{2\widetilde m + 1}}{2\widetilde m + 1}.
$$

Equation (\ref{wms}), does not lead in this case to a simple expression for 
$\widetilde m_s$, as it did in Example 3. We can nevertheless still derive very
useful information from it. To simplify the resulting equation, set 
$R = \widetilde R^0_s = 1/(2\widetilde m_s + 1)$. Then (\ref{wms}) reads
\be
C \, - \, B R \ = \ (T-1) \, \left\{ BR \, + \, 
\frac{B \widetilde a - C}{1-\widetilde a}
\right\} \, (1-\widetilde a)^{1/R},
\label{example2R}
\ee
except in the trivial case $\widetilde a = 1$, in which the right hand side of (\ref{example2R})
is 0. This case corresponds to $a = \widetilde a \, n = n$, so that types A cooperate only in 
the first round. Obviously then, Example 2 reduces to Example 1, and indeed, 
(\ref{example2R}) reduces to (\ref{HRpgws}).

In the opposite extreme, when $\widetilde a = 0$, so that $a = \widetilde a \, n = 0$ and 
types A cooperate in each round of the game, the right hand side of (\ref{example2R})
reduces to $(T-1) \, (C-BR)$. So (\ref{example2R}) is again reduced to (\ref{HRpgws}), 
as one should expect. 

Note that also when $T=1$, Example 2 reduces to Example 1, and 
(\ref{example2R}) reduces to Hamilton's rule (\ref{HRpgws}).
One can see the right hand sight of (\ref{example2R}), as
a correction to that form of the viability condition in the more general Example 2.


Fig. \ref{fig15} compares the exact values of $m_s$ and $R^0_s$ for instances of 
Example 2, under weak selection, from 
the viability condition (\ref{HR+wspi}), and the approximation provided by solving
(\ref{example2R}). Notice in these graphs, that $R^0_s$ is significantly smaller 
than $C/B$, when $\widetilde a$ is close to $C/B$.  

While somewhat intimidating at first sight, (\ref{example2R}) provides good information about
the critical value $\widetilde R^0_s$. We first state the main features of its behavior, and
then explain how to do the computations leading to these claims. 
Of special importance is the case 
in which $\widetilde a = C/B = \min \{0 \leq x \leq 1 : \widetilde v^A_x \geq 0 \}$. 
In this case (\ref{example2R}) simplifies to:
\be
C \, - \, BR \ = \ (T-1) \, BR \, (1-C/B)^{1/R}.
\label{example2RaCB}
\ee
With $C$, $B$ and $T$ fixed, $\widetilde R^0_s$ is a continuous function of $0 \leq \widetilde a \leq 1$,
that takes the value $C/B$ on both end-points of this domain, is strictly decreasing when 
$0 \leq \widetilde a \leq C/B$, and strictly increasing on $C/B \leq \widetilde a \leq 1$. 
It reaches therefore its minimum at $\widetilde a = C/B$, where it solves (\ref{example2RaCB}). 
It is not surprising that when $C/B \leq \widetilde a < 1$, so that (\ref{mtft}) is
satisfied, we have $\widetilde R^0_s < C/B$, since in this case, when altruists
continue cooperating, it is always in their interest to do so.
But it is somewhat surprising that also when $0 < \widetilde a < C/B$, we have
$\widetilde R^0_s < C/B$. It is intuitive that $\widetilde R^0_s$
should reach its minimal value when $\widetilde a = C/B$, since then
altruists are persisting precisely when they should.

With $T$ and $\widetilde a$ fixed, $\widetilde R^0_s$ is a decreasing function of 
$B/C$, that goes to 0 as $B/C \to \infty$ (Fig. \ref{fig17}).
With $C$, $B$ and $0 < \widetilde a < 1$ fixed, $\widetilde R^0_s$ is a decreasing function of $T$ that 
behaves as follows when $T \to \infty$. If $0 \leq \widetilde a < C/B$, then
$\widetilde R^0_s \to (C - B \widetilde a) / (B - B \widetilde a)$, while if 
$C/B \leq \widetilde a \leq 1$, then $\widetilde R^0_s \to 0$.
In this latter case, the convergence is rather slow, in that 
$\widetilde R^0_s$ behaves asymptotically as $(\log(1/(1-\widetilde a)))/\log (T)$.
(See Fig. \ref{fig18}.)

The claims above follow from the behavior of the left hand side and the right hand side of 
(\ref{example2R}). We denote them respectively by $H(R) = H_{C,B}(R)$ and $G(R) = G_{C,B,T,\widetilde a}(R)$.
(With $G(0) = 0$, so that $G$ is continuous on the interval $[0,1]$. 
Note that not only $G(R)$, but 
also all its derivatives converge to 0 as $R \to 0$.) See Fig. \ref{fig19}, for an illustration 
of what follows. 
The function $H(R)$ is very straightforward; it is a strictly decreasing function of $R$,
that is positive for $R < C/B$ and negative for $R > C/B$. 
The function $G(R)$ has the sign of the
term inside the curly braces. That term inside the curly braces is 0 when
$R = (C-B\widetilde a)/(B-B\widetilde a)$. We define
$\widehat R = \max\{(C-B\widetilde a)/(B-B\widetilde a),0\}$, and
observe that $0 < \widehat R < C/B$, when $\widetilde a < C/B$,
and $\widehat R = 0$, when $\widetilde a \geq C/B$.
The term inside the curly braces, and therefore also $G(R)$, is negative for
$R < \widehat R$, and positive for $R > \widehat R$. Notice also that once it
is positive, $G(R)$ is strictly increasing in $R$, since it is then the product of
strictly increasing positive functions. The behaviors described so far for $H(R)$ and $G(R)$
immediately imply that they are equal to each other at exactly one
point $R = \widetilde R^0_s$, and that this point is in the open interval
$(\widehat R,C/B)$.

The various claims about the behavior of $\widetilde R^0_s$ as a function of 
$\widetilde a$, or of $B/C$, or of $T$, follow now from analyzing the behavior 
of the graphs of $H(R)$ and of $G(R)$, as these parameters change. 
We have
$\partial G_{C,B,T,\widetilde a}(R)/\partial \widetilde a  \, = \, 
(T-1) \, (C - B \widetilde a) \, ((1/R) - 1) \, (1-\widetilde a)^{(1/R) -2}$,
which, regardless of the value of $R$,
is positive for $0 < \widetilde a < C/B$ and negative for $C/B < \widetilde a < 1$.
This means that the graph of $G(R)$ moves upwards, as $\widetilde a$ increases 
from 0 to $C/B$, and then moves downwards, as $\widetilde a$ increases 
from $C/B$ to 1. In the extremes, $G(R) \to (T-1) (BR - C)$, as 
$\widetilde a \to 0$, for all $R > 0$. (This convergence is not uniform,
since the function $G(R)$ 
converges to 0 as $R \to 0$.)
And $G(R) \to 0$, uniformly in $R$, as $\widetilde a \to 1$.
These facts, and the trivial behavior of $H(R)$ that does not depend on 
$\widetilde a$, provides us with the facts about the dependence of 
$\widetilde R^0_s$ on $\widetilde a$. 

The fact that $\widetilde R^0_s$
depends on $C$ and $B$ only through $B/C$ can be seen by dividing both sides of
(\ref{example2R}) by $C$. 
The fact that $\widetilde R^0_s$ decreases as $B/C$ increases follows easily from observing that 
the graphs of $H(R)$ and $G(R)$ move, respectively, down and up, as $B$ increases, with $C$ 
fixed. 
And the fact that $\widetilde R^0_s \to 0$,
as $B/C \to \infty$, is immediate from $0 < \widetilde R^0_s < C/B$ (Fig. \ref{fig17}). 

If we keep $B$, $C$ and $0 < \widetilde a < 1$ fixed and let
$T \nearrow \infty$, then $G(R)$ also goes monotonically to $\infty$, for $\widehat R < R < C/B$,
and stays at 0 for $R = \widehat R$. The corresponding behavior of the graph of 
$G(R)$, and the fact that $\widetilde R^0_s$ is the point in $\widehat R < R < C/B$
where this graph intersects the graph of the decreasing function $H(R)$, shows that,
as claimed above,  
$\widetilde R^0_s$ is decreasing in $T$, and as $T \to \infty$, 
$\widetilde R^0_s \to \widehat R$ (Fig. \ref{fig20}). 
The claim about the slow speed of this convergence, in case $C/B \leq \widetilde a < 1$,
can be obtained by taking the logarithm of both sides of (\ref{example2R}) and 
analyzing how the resulting terms behave as $T \to \infty$.

\

It is interesting to contrast (\ref{ipgR0s}), for iterated pairwise prisoner dilemma with 
types N defecting and types A playing tit-for-tat 
(Example 4, (\ref{ipd}))
with our analysis above of the 
behavior of $R^0_s$ in Example 2, which is an iterated public goods 
game (an analogue of the prisoner dilemma in a multi-individual setting), 
with types N defecting and types A playing many-individual tit-for-tat.
There are several expected similarities in the behavior of $R^0_s$ in both 
cases, as a function of costs, benefits and expected number of repetitions of
the game. But there are also important differences to emphasize. 
There are differences in the details of the behavior. For instance,
$R^0_s$ goes to 0, as $T \to \infty$, much more slowly 
in the case of Example 2, when it does go to 0. But equally important, 
we want to mention 
the differences in the level of complexity of the analysis in each case. 
While (\ref{ipgR0s}) holds for arbitrary $n$ and resulted from a standard 
Hamilton rule, (\ref{HRpgws}), our analysis of Example 2 above relied on the much 
more elaborate results developed in this section, and depended on $n$ being large.
One of the main messages of the current paper is that when interactions 
involve several individuals at a time in the groups, one needs methods that
go beyond those that apply to pairwise additive interactions. 

\

It is important to explain why the assumptions made in \cite{BR} on how 
much assortment to expect in Example 2, and later used in several papers,
including \cite{BGB}, are excessively pessimistic. In \cite{BR} it 
was supposed that conditioned on the focal individual being type A, the
other $n-1$ members of the focal group would be type A independently, 
with probability $\P_t(A_2|A_1)$. In our framework, given that the focal
is type A, and that the co-focal is also type A, further increases the 
conditional probability that a third member of the group is type A. 
Given then that a third individual is type A, again further increases
the probability that a fourth individual is type A, and so on. This is so 
because the information being successively provided keeps increasing the 
probability that there were several types A in the previous generation 
in the group from which the focal descends. 
It is very interesting to make the computation
of $\widetilde R^0_s$ using the assumption of \cite{BR} and compare the result 
with what we have obtained from 
(\ref{example2R}). Under that assumption of conditional independence, 
the viability condition (\ref{HR+wspi}) for the regime of weak selection,
would be replaced with $\sum_k \, v^A_k \, \P(\mbox{Bin}(n-1,R^0)=k-1) = 0$.
Assume that (\ref{limv}) holds and define 
$\widetilde v^{A*}_x$ as $\widetilde v^{A}_x$, when $x$ is a continuity point of 
$\widetilde v^{A}_x$, and as the average between the limits of 
$\widetilde v^{A}_x$ from the left and from the right at $x$, otherwise. 
(We are supposing that these limits exist, as is the case in Example 2.)
Then, by a central limit theorem, in the limit (\ref{nmlimit}).
$$
\sum_k \, v^A_k \, \P(\mbox{Bin}(n-1,R^0)=k-1) \ \ \longrightarrow \ \ 
\widetilde v^{A*}_{\widetilde R^0}.
$$
(Compare with (\ref{wVA}).) Instead of (\ref{wms}), we would then have 
\be
\widetilde v^{A*}_{\widetilde R^0_s} \ = \ 0.
\label{wmsbr}
\ee
For Example 2, the only solution of (\ref{wmsbr}) is $\widetilde R^0_s = C/B$,
regardless of the value of $\widetilde a$. This result would not depend on $T$,
and would differ substantially from our result in case, e.g., 
$\widetilde a \geq C/B$, $T$ large, which has $\widetilde R^0_s < < C/B$.
(See Fig. \ref{fig15}, Fig. \ref{fig17}, Fig. \ref{fig18} and Fig. \ref{fig20}.)



\clearpage

\begin{figure}[th!]
\begin{center}
\includegraphics[width=12cm]{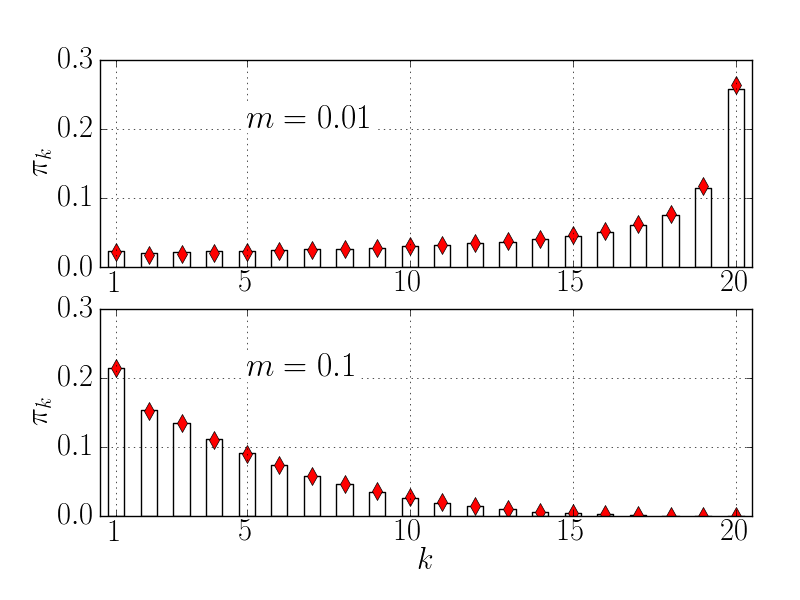}
\end{center}
\caption{Distribution $\pi_k$ given by (\ref{pinu}), or (\ref{piQ}) (bars) compared with 
 $k\nu_k^{\delta}/\sum_{k^\prime} k^{\prime}\nu_{k^\prime}^{\delta}$, where $\nu_k^\delta$ is the Perron-Frobenius 
eigenvector  with $\delta=0.01$  for the Threshold model (THR, Example 3)  with parameters $n=20$,
 $C=1$, $A=A^\prime=5$ and $\theta=4$ (red diamonds). The comparison is repeated for migration 
rates  $m=0.01$ (top panel) and $m=0.1$ (bottom panel). } 
\label{fig11}
\end{figure}

\begin{figure}[th!]
\begin{center}
\includegraphics[width=12cm]{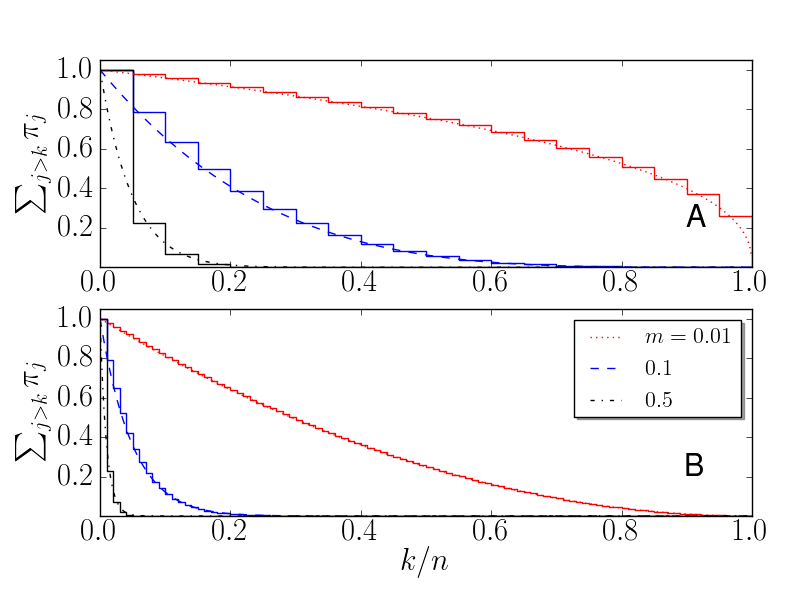}
\end{center}
\caption{Limit of large $n$ and small $m$ under weak selection. This figure compares tail probabilities for the
 distribution $\pi_k$ provided by (\ref{pinu}) (stairs) and for Beta densities with parameters
 $\alpha=1$ and $\beta=2\tilde{m}$. Panel A shows the case $n=20$ for, from top to bottom, 
$m=0.01$ (red dotted line), $m=0.1$ (blue dashed line) and $m=0.5$ (black dot dashed line). Panel B 
depicts the same scenarios for the case $n=100$.}
\label{fig12}
\end{figure}

\begin{figure}[th!]
\begin{center}
\includegraphics[width=12cm]{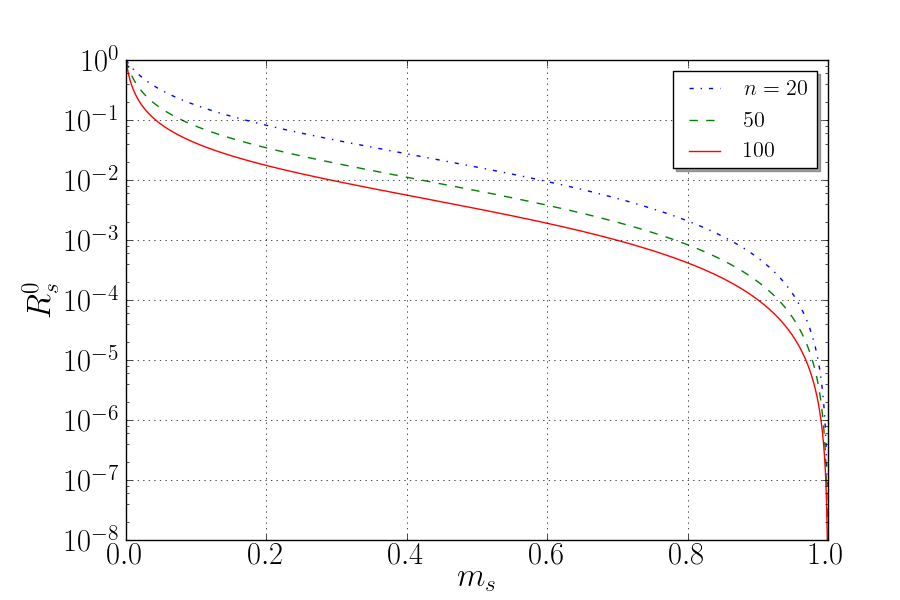}
\end{center}
\caption{Relatedness and migration rate under weak selection. For ease of comparison this figure depicts the 
relatedness $R_s^0$ (\ref{R0s=R0s(ms)}) as a function of the migration rate $m_s$ for, from top to bottom, $n=20$ (dot-dashed blue line),
$n=50$ (dashed green line) and $n=100$ (full red line).}
\label{fig13}
\end{figure}

\begin{figure}[th!]
\begin{center}
\includegraphics[width=12cm]{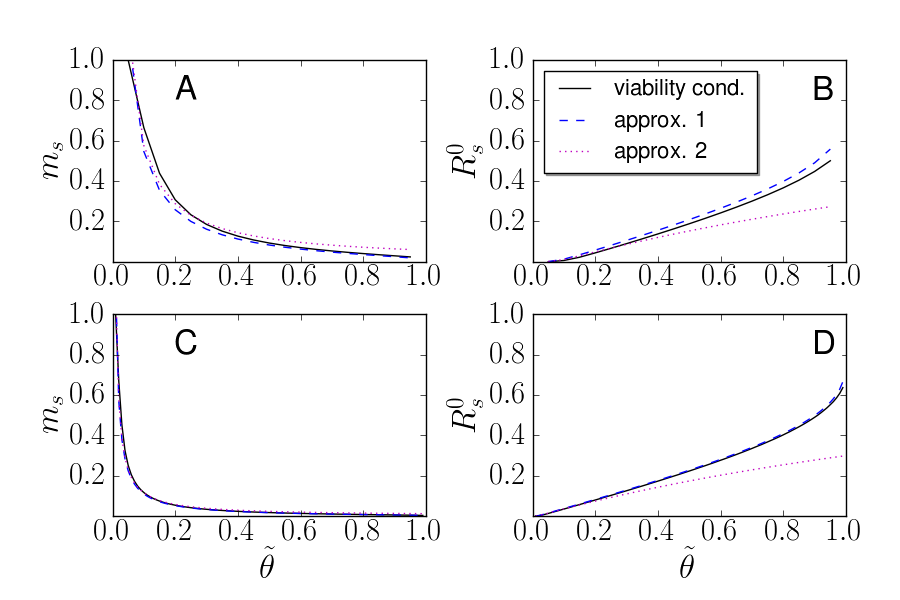}
\end{center}
\caption{Limit of large $n$ and small $m$ under weak selection for the Threshold model (Example 3). Panels represent 
critical migration rates (A and C) and critical relatedness (B and D)  
for the Threshold model with $C=1$ and $A=A^{\prime}=10$ as a function of $\tilde{\theta}=\theta/n$.
Top panels A and B depict the case $n=20$. Bottom panels C and D depict the case $n=100$. In each panel critical values 
obtained  by the viability condition under weak selection (\ref{HR+wspi}), or (\ref{HRtmws}) (viability cond., black full lines) are compared with the 
approximation for large $n$ and small $m$ given by (\ref{msapproxex3}) (approx. 1, dashed blue lines) and with the approximation 
(\ref{msapproxex3'}) that assumes $n$ large, $m$ small and  also $\tilde{\theta} \ll 1$ (approx. 2, dotted red lines).}
\label{fig14}
\end{figure}

\begin{figure}[th!]
\begin{center}
\includegraphics[width=12cm]{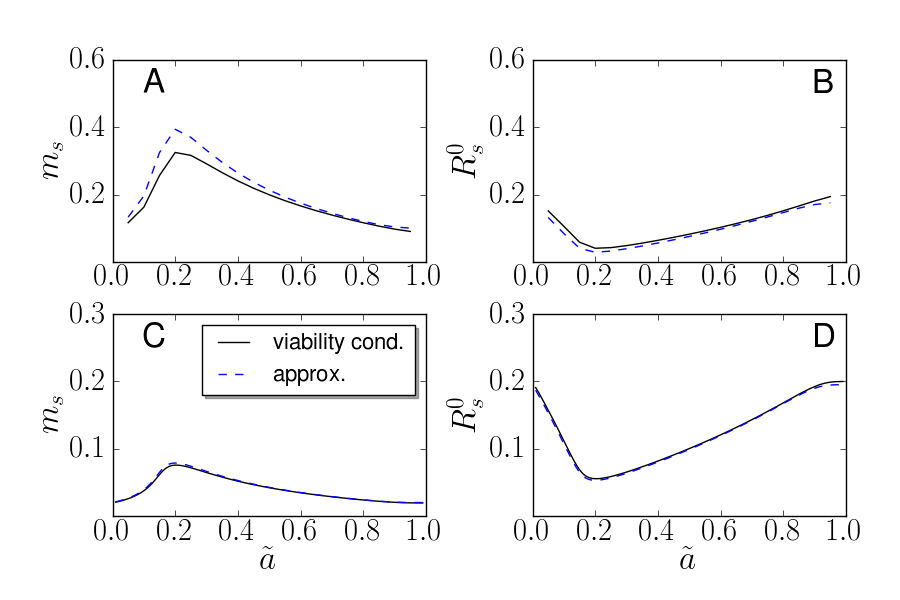}
\end{center}
\caption{Limit of large $n$ and small $m$ under weak selection for the  Iterated public goods (IPG) game (Example 2). 
Panels represent critical migration rates (A and C) and critical relatedness (B and D) for the IPG  with $C=1$, $B=5$ 
and $T=100$ as a function of $\tilde{a}=a/n$. Top panels A and B depict the case $n=20$. Bottom panels C and D depict the case $n=100$. 
In each panel critical values obtained  by the viability condition under weak selection (\ref{HR+wspi}) 
(viability cond., black full lines) are compared with the 
approximation for large $n$ and small $m$ given by solving  (\ref{example2R}) in $R$ (approx., dashed blue lines).
In panel B we have $R^0_s=4.02\%$ when $\tilde{a}=20\%$, and in panel D we have $R^0_s=5.54\%$ when $\tilde{a}=20\%$.}
\label{fig15}
\end{figure}

\begin{figure}[th!]
\begin{center}
\includegraphics[width=12cm]{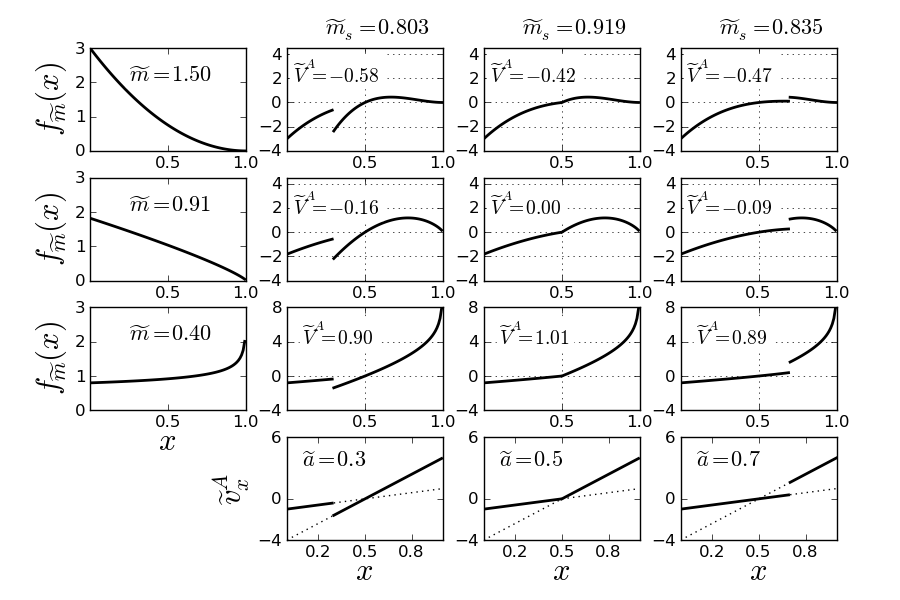}
\end{center}
\caption{Viability condition in the limit of large $n$ and small $m$ under weak selection. The case illustrated is 
the  Iterated public goods (IPG) game (Example 2) with with $C=1$, $B=2$ and $T=4$. The first column
 depicts the density $f_{\widetilde{m}}(x)$ for $\widetilde{m}=1.50,0.91,0.40$. The bottom row represents the payoff function 
$\widetilde{v}_x^A$ for $\widetilde{a}=0.3,0.5,0.7$. The grid with nine plots represents  $\widetilde{v}_x^A f_{\widetilde{m}}(x)$ 
with  $\widetilde{m}$ and $\widetilde{a}$ as specified in each row and column. The viability condition is given by 
   $\widetilde{V}^A(\widetilde{m})=\int_0^1 dx\,\widetilde{v}_x^A f_{\widetilde{m}}(x) >0$. 
As $\widetilde{m}$ is decreased, the positive part of the integrand increases, eventually reaching the critical value 
$\widetilde{m}_s$ (annotated in the top of each column). 
Payoffs are maximized for $\tilde{a}=C/B$, implying a maximal value for 
$\widetilde{m}_s$ (in the case depicted $\widetilde{m}_s=0.919$).  Decreasing  $\widetilde{a}$, 
increases the negative part of $\tilde{v}_x^A f_{\tilde{m}}(x)$ and, consequently, decreases $\widetilde{m}_s$ 
(increases $\widetilde{R}^0_s$). Increasing $\widetilde{a}$, decreases the positive part of the integrand and 
also decreases $\widetilde{m}_s$.}
\label{fig16}
\end{figure}

\begin{figure}[th!]
\begin{center}
\includegraphics[width=12cm]{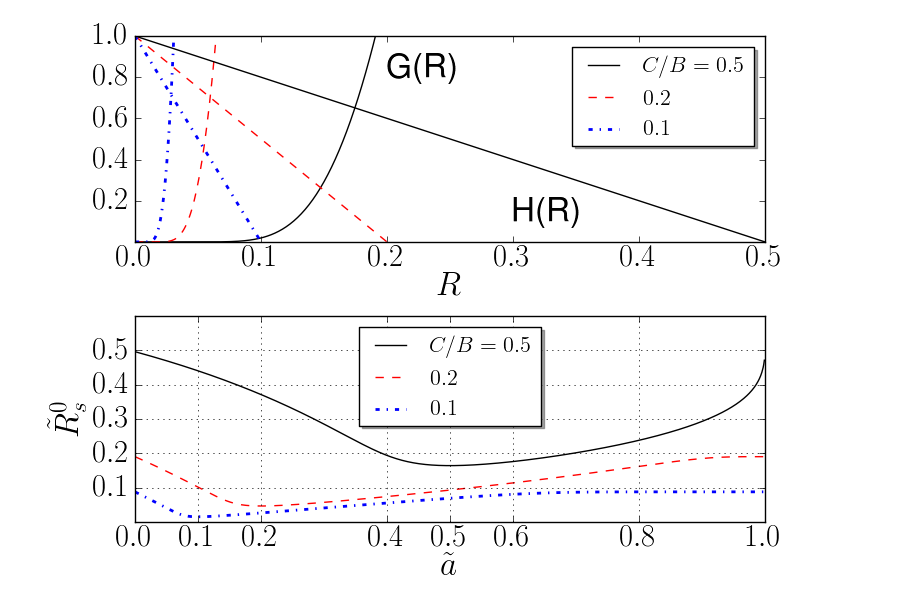}
\end{center}
\caption{Limit of large $n$ and small $m$ under weak selection for the  Iterated public goods (IPG) game  (Example 2): 
behavior of solutions for (\ref{example2R}) - Part 1. Top panel: $H(R)$ corresponds to the l.h.s. of  (\ref{example2R}) while $G(R)$ 
depicts the r.h.s. of (\ref{example2R}). $H(R)$ is strictly decreasing and it is positive for $R<C/B$. Derivatives of $G(R)$ 
converge to $0$ as $R\rightarrow 0$. $H(R)$ and $G(R)$ are equal to each other at exactly one point $R=\tilde{R}_s^0$ that 
is a decreasing function of $C/B$. Curves depicted correspond to the cases $C/B=0.5$ (full black line), $C/B=0.2$ (dashed red
line) and $C/B=0.1$ (dot-dashed blue line) with $\tilde{a}=C/B$ and $T=100$. Bottom panel: $\tilde{R}_s^0$ as a function
of $\tilde{a}$ for $C/B=0.5$ (top, full black line), $C/B=0.2$ (middle, dashed red line) and $C/B=0.1$ (bottom, dot-dashed 
blue line) and $T=100$. $\tilde{R}_s^0$ is continuous  in the interval $0\le \tilde{a} \le 1$,  takes the 
value $C/B$ on both end-points of this domain and has a minimum at $\tilde{a}=C/B$. }
\label{fig17}
\end{figure}

\clearpage
\begin{figure}[th!]
\begin{center}
\includegraphics[width=12cm]{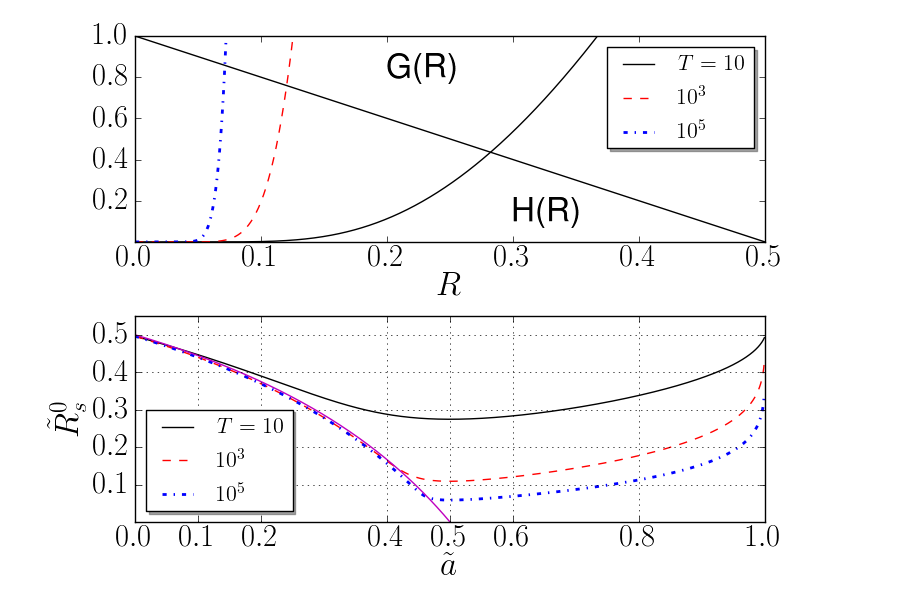}
\end{center}
\caption{Limit of large $n$ and small $m$ under weak selection for the  Iterated public goods (IPG) game  (Example 2): 
behavior of solutions for (\ref{example2R}) - Part 2. Top panel: $G(R)$ and $H(R)$ for  $C/B=0.5$, $\tilde{a}=0.5$  and 
$T=10$ (leftmost, full black line), $T=10^3$ (dashed red line) and $T=10^5$ (dot-dashed blue line). $\tilde{R}_s^0$ is a 
decreasing function of $T$. Bottom panel: in the limit $T\rightarrow\infty$, if $0\le \tilde{a}<C/B$ then 
 $\tilde{R}_s^0 \rightarrow \frac{C/B -\tilde{a}}{1-\tilde{a}}$ (full magenta line). If $C/B\le \tilde{a}\le 1$ then 
$\tilde{R}_s^0 \rightarrow 0$ very slowly.}
\label{fig18}
\end{figure}

\begin{figure}[th!]
\begin{center}
\includegraphics[width=12cm]{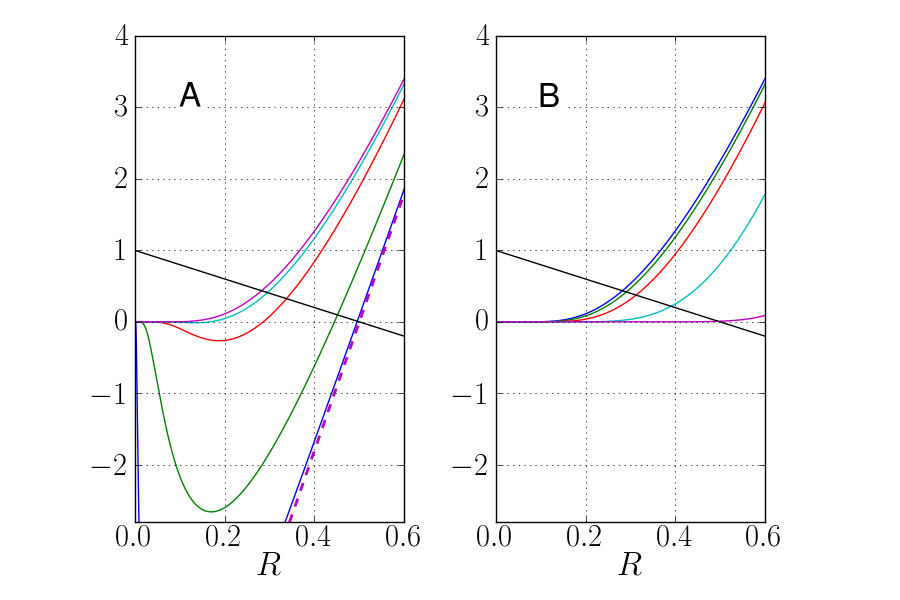}
\end{center}
\caption{Limit of large $n$ and small $m$ under weak selection for the  Iterated public goods (IPG) game  (Example 2): 
behavior of solutions for (\ref{example2R}) - Part 3. $H(R)$ (strictly decreasing straight line) and $G(R)$ for $C/B=0.5$ and 
$T=10$ for $\tilde{a}=0.01,0.1,0.3,0.4,0.5$ from right to left in Panel A and for $\tilde{a}=0.5,0.6,0.7,0.9,0.999$ from left to
right in Panel B. The graph of $G(R)$ moves upwards for $0\le \tilde{a}<C/B$ and downwards for $C/B\le \tilde{a}\le 1$. 
$G(R)\rightarrow (T-1)(BR-C)$ as $\tilde{a} \rightarrow 0$ (dashed magenta line in Panel A). In Panel B it can be seen that
$G(R)\rightarrow 0$ as $\tilde{a} \rightarrow 1$.   }
\label{fig19}
\end{figure}

\begin{figure}[th!]
\begin{center}
\includegraphics[width=12cm]{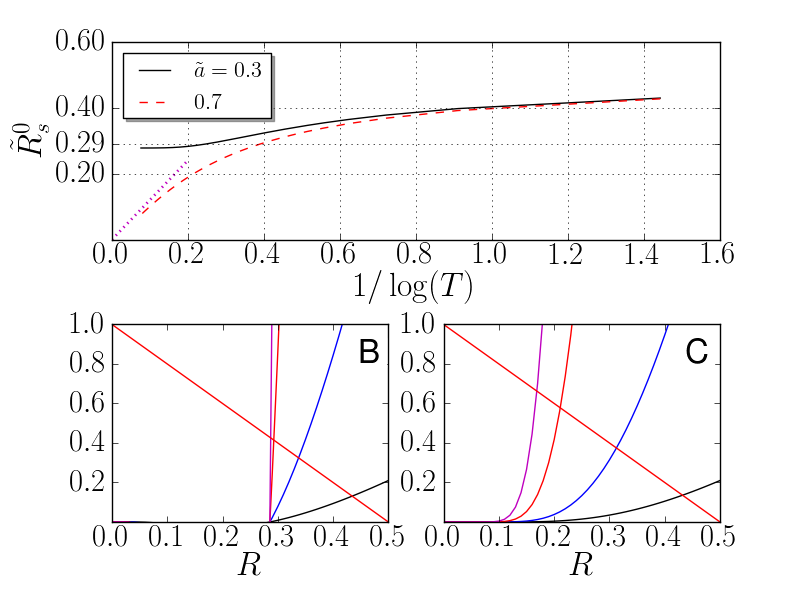}
\end{center}
\caption{Limit of large $n$ and small $m$ under weak selection for the  Iterated public goods (IPG) game  (Example 2): 
behavior of solutions for (\ref{example2R}) - Part 4. In all panels $C/B=0.5$. Panel A depicts $\tilde{R}_s^0$ as a 
function of $1/\log(T)$ for $\tilde{a}=0.3$ (full black line) and for $\tilde{a}=0.7$ (dashed red line). 
For $0\le \tilde{a}<C/B$  $\tilde{R}_s^0 \rightarrow \frac{C/B -\tilde{a}}{1-\tilde{a}}$ 
(this value is approximately $0.286$ for the case shown). If $C/B\le \tilde{a}\le 1$ then $\tilde{R}_s^0$ converges to $0$ very 
slowly as $T$ increases, more specifically $\tilde{R}_s^0 \sim -\log(1-\tilde{a})/\log(T)$ (dotted magenta line). Bottom panels 
show the behavior of $G(R)$ as $T$ increases. Panel B: case $\tilde{a}<C/B$ for, from right to left, $T=2,10,100,500$. Panel A: 
case $\tilde{a}>C/B$ for $T=2,10,100,500$, from right to left. $G(R)$ stays at zero for 
$\hat{R}=\max\{\frac{C/B -\tilde{a}}{1-\tilde{a}},0\}$ and goes monotonically to infinity for $\hat{R}<R<C/B$.}
\label{fig20}
\end{figure}

\end{document}